\documentclass[aps,prd,twocolumn,nofootinbib,floatfix]{revtex4-1}
\usepackage{setspace}
\usepackage{amsthm}
\theoremstyle{definition}

\usepackage{graphicx}
\usepackage{dcolumn}
\usepackage{multirow}
\usepackage{subfigure}
\usepackage{times,mathptm}
\usepackage{float}
\usepackage{color}
\usepackage{amsmath,amsfonts}
\usepackage{mathptmx}
\usepackage{mathrsfs}
\usepackage{bbm}
\usepackage{bm}
\usepackage{xfrac}

\newcommand{\beq}{\begin{equation}}
\newcommand{\eeq}{\end{equation}} 
\newcommand{\bea}{\begin{eqnarray}}
\newcommand{\eea}{\end{eqnarray}}

\newcommand{\V}{\mathcal{V}}

\newcommand{\C}{{\cal C}}

\renewcommand{\d}{\delta}

\renewcommand{\l}{\lambda}
\renewcommand{\L}{\Lambda}

\newcommand{\F}{{\cal F}}

\newcommand{\tb}{\tilde{\beta}}
\newcommand{\tg}{\tilde{\gamma}}
\newcommand{\p}{\phi}
\renewcommand{\b}{\beta}
\renewcommand{\a}{\alpha}

\renewcommand{\ni}{\noindent}
\newcommand{\tr}{\text{Tr}}

\newcommand{\vx}{{\vec{x}}}
\newcommand{\vy}{{\vec{y}}}
\newcommand{\vz}{\vec{z}}

\newcommand{\n}{\nu}
\newcommand{\m}{\mu}

\newcommand{\g}{\gamma}

\newcommand{\s}{\sigma}
\renewcommand{\k}{\kappa}
\newcommand{\D}{\Delta}

\newcommand{\G}{\Gamma}

\newcommand{\N}{{\cal N}}

\newcommand{\M}{{\cal M}}
\renewcommand{\th}{\theta}

\newcommand{\vph}{\varphi}
\newcommand{\oh}{\frac{1}{2}}

\newcommand{\dg}{\dagger}
\newcommand{\non}{\nonumber}

\newcommand{\rf}[1]{(\ref{#1})}
\newcommand{\ra}{\rightarrow}
\newcommand{\pa}{\partial}
\renewcommand{\vec}[1]{\bm #1}

\usepackage{ulem}

\begin{document}

\title{What symmetry is actually broken in the Higgs phase of a gauge-Higgs theory?} 

\bigskip
\bigskip

\author{Jeff Greensite and Kazue Matsuyama}
\affiliation{Physics and Astronomy Department, San Francisco State
University,   \\ San Francisco, CA~94132, USA}
\bigskip
\date{\today}
\vspace{60pt}
\begin{abstract}

\singlespacing
 
    In SU($N$) gauge-Higgs theories, with a single Higgs field in the fundamental representation, there exists in addition to the local gauge symmetry a global SU(2) symmetry, at $N=2$, and a global U(1) symmetry, for $N \ne 2$.     We construct a gauge-invariant 
order parameter for the breaking of these global symmetries in the Higgs sector, and calculate numerically the transition lines, in coupling-constant space, for SU(2) and SU(3) gauge theories with unimodular Higgs fields.  The order parameter is non-local, and therefore its non-analyticity does not
violate the theorem proved by Osterwalder and Seiler.  We then show that there exists a transition, in gauge-Higgs theories, between two types of confinement: ordinary color neutrality in the Higgs region, and a stronger  condition, which we have called ``separation-of-charge confinement," in the confinement region.  We conjecture that the symmetry-breaking transition coincides with the transition between these
two physically different types of confinement.

\end{abstract}

\pacs{11.15.Ha, 12.38.Aw}
\keywords{Confinement,lattice
  gauge theories}
\maketitle

\singlespacing
\section{\label{intro}Introduction}

   Contrary to statements found in some textbooks, a local gauge symmetry cannot be broken spontaneously, as shown long ago
by Elitzur  \cite{Elitzur:1975im}.
In certain gauges there are remnant global symmetries which {\it can} break spontaneously,
but the locations of the corresponding transition lines are gauge dependent \cite{Caudy:2007sf}, which makes a physical interpretation of such
transitions dubious.  Of course, in a gauge-Higgs theory with the scalar field in the fundamental representation, there is a
confinement-like region analogous to QCD, in which one finds color electric flux tube formation, Regge trajectories, and a linear
static quark potential followed by string breaking.  There is also a Higgs region with no flux tube formation, no Regge trajectories,
and only Yukawa forces between static sources.  It was shown many years ago by Osterwalder and Seiler
\cite{Osterwalder:1977pc}, whose work was further elucidated by Fradkin and Shenker \cite{Fradkin:1978dv}, that there is no
thermodynamic transition which entirely isolates the Higgs regime from the confinement-like regime, meaning that the free energy
is analytic along some path between any two points in coupling constant space.  The implication is that, in the absence of a massless phase, there can be no transition from a color neutral to a color charged spectrum of asymptotic states.  The gauge-invariant composite operators which create color-neutral physical particles in the electroweak theory were first written down by Fr\"ohlich, Morchio, and Strocchi 
\cite{Frohlich:1981yi,*Frohlich:1980gj} and by 't Hooft \cite{tHooft:1979yoe}.

    All these facts appear to imply that, in the absence of a gauge choice, there is no such thing as spontaneous symmetry
breaking in the context of the Brout-Englert-Higgs (BEH) mechanism, and no gauge-invariant order parameter which could detect
such a breaking.  On the other hand, it is well known that in SU(2) gauge-Higgs theory there exists a global SU(2) symmetry, distinct from the local gauge symmetry, and it was likewise pointed out by Maas et al.\ \cite{Maas:2016ngo,*Maas:2017xzh,*Maas:2017pcw} that for SU($N>2$) gauge-Higgs theories the additional symmetry is
global U(1).  But although global symmetries can break spontaneously, the absence of massless Goldstone excitations would seem to rule
out that possibility.  In this article we point out that global symmetries in the Higgs sector can break in the Higgs sector, in the sense explained below, without introducing
Goldstone particles in the full theory.  We will construct a gauge-invariant order parameter which is sensitive to these symmetry breakings, and
map out the transition line in coupling constant space for SU(2) and SU(3) gauge-Higgs theories with a single unimodular Higgs
field. 

     This raises the question of the physical distinction between the symmetric and broken phases of a gauge-Higgs theory.  In a recent
article \cite{Greensite:2017ajx}  we have suggested that gauge theories with matter fields in the fundamental representation may satisfy a confinement criterion which is stronger than the usual condition of a color-neutral spectrum.  This stronger condition, which
is a generalization of the Wilson area law criterion to gauge + matter theories,  is called ``separation of charge'' or $S_c$ confinement, although its existence beyond pure gauge theory was only conjectured in ref.\ \cite{Greensite:2017ajx}.  In the present article, we show
that $S_c$ confinement actually  does exist in at least some region of the gauge-Higgs phase diagram, which implies
the existence of a transition between the stronger and weaker confinement phases.  It is therefore natural to suppose, although we do not prove, that the symmetry-breaking transition in gauge-Higgs theories, which we have located here for the SU(2) and SU(3) gauge groups, corresponds to a transition between these two, physically distinct, types of confinement.

    We should note that other criteria for confinement with matter fields can be found in the literature, namely the Kugo-Ojima
criterion \cite{Kugo:1979gm}, non-positivity or unphysical pole structure in quark/gluon propagators (an early reference is \cite{Stingl:1985hx}), and the Fredenhagen-Marcu proposal \cite{Fredenhagen:1985ft}.  The first two of these proposals rely on BRST symmetry, which is dubious at the non-perturbative level, while the Fredenhagen-Marcu criterion only distinguishes between massless and massive phases, rather than between Higgs and confinement.  For a more detailed critique, see section V of \cite{Greensite:2017ajx}.

\section{\label{sec2}SU(2) gauge-Higgs theory}

    The symmetry of an SU(2) gauge-Higgs theory with a single Higgs doublet is SU(2)${}_{gauge} \times$ SU(2)${}_{global}$.  The
extra global symmetry is easiest to see by mapping the Higgs doublet (which we take, for simplicity, to be unimodular $|\vec{\phi}|=1$),
onto an SU(2) group element
\bea
            \vec{\phi} =\left[ \begin{array}{c} \phi_1 \cr \phi_2  \end{array} \right] \Longrightarrow
                  \phi = \left[ \begin{array}{cc}
                                               \phi_2^* & \phi_1  \cr
                                               -\phi_1^*  & \phi_2  \end{array} \right]  \ ,
\label{map}
\eea
and the action can then be written in the form
\bea
     S &=& S_W[U] + S_H[\phi,U]  \non \\
        &=& - \beta \sum_{plaq} \oh \mbox{Tr}[U_\m(x)U_\n(x+\hat{\m})U_\m^\dg(x+\hat{\n}) U^\dg_\n(x)]  \non \\
   & &       - \gamma \sum_{x,\m} \oh \mbox{Tr}[\phi^\dg(x) U_\m(x) \phi(x+\widehat{\m})] \ ,
\label{Sgh}
\eea
which has the following invariance
\bea
            U_\m(x) &\ra& L(x)  U_\m(x) L^\dg(x+\hat{\mu})  \non \\
            \phi(x) &\ra& L(x) \phi(x) R \ ,
\eea
where $L(x) \in $ SU(2)${}_{gauge}$ is a local gauge transformation, while $R \in $ SU(2)${}_{global}$ is a global transformation
                     
      If we choose a gauge (e.g.\ unitary gauge) in which the Higgs field acquires a vacuum expectation value (VEV)
\beq
    \langle \phi \rangle = \left[ \begin{array}{cc}
                                               \upsilon & 0 \cr
                                               0 & \upsilon \end{array} \right] \ ,
\eeq      
then the SU(2)${}_{gauge} \times$ SU(2)${}_{global}$ symmetry is broken down to a diagonal global subgroup
\beq
\mbox{SU(2)}_{gauge} \times \mbox{SU(2)}_{global} \ra \mbox{SU(2)}_D \ ,
\eeq
corresponding to transformations
\bea
           L(x) &=& R^\dg =G  \non \\
           \phi(x) &\ra& G \phi(x) G^\dg  ~~,~~ U_\m(x) \ra G U_\m(x) G^\dg \ .
\eea
Transformations in this diagonal subgroup, which is known as the group of ``custodial symmetry,'' preserve the VEV of $\phi$.
Custodial symmetry has a role to play in the phenomenology of the electroweak interactions, and 
is reviewed in many places, e.g.\ \cite{Willenbrock:2004hu,Weinberg:1996kr,Maas:2017wzi}.
Here, however, we would like to focus on the $R$-transformations belonging to SU(2)${}_{global}$.\footnote{The term ``custodial symmetry''
is sometimes used to refer to the group SU(2)${}_{global}$ of $R$-transformations, rather than the diagonal subgroup SU(2)$_D$.  See, e.g.,
Maas \cite{Maas:2017wzi}.}

    We have already noted that characterizing the Brout-Englert-Higgs mechanism as a spontaneous breaking of gauge symmetry due to the non-zero VEV of $\phi$ is rather misleading, given that
\begin{itemize}
\item  $\langle \phi \rangle =0$ at all $\b, \g$ in the absence of a gauge choice;
\item $\langle \phi \rangle  \ne 0$ at all $\b,\g$ in unitary gauge;\footnote{To this we might add that the lattice abelian-Higgs model in four dimensions has a massless phase in some region of the $\b-\g$ plane \cite{Ranft:1982hf,Fradkin:1978dv}, despite the fact that in unitary gauge $\langle \phi \rangle \ne 0$ also in that region.} 
\item In other gauges $\langle \phi \rangle$ may be zero or non-zero at a given $\b,\g$, depending on the gauge choice.
\end{itemize}

   But if the VEV of $\phi$ is misleading, at least outside the context of perturbation theory, we may still ask whether the Higgs phase
of an SU(2) gauge-Higgs theory can be distinguished from a non-Higgs phase by the spontaneous breaking of the SU(2)${}_{global}$ symmetry. This question is motivated by the fact that a non-zero (but gauge-dependent) $\langle \phi \rangle$ always implies a broken SU(2)${}_{global}$.  The idea is to turn this around, i.e.\ the signature for the Higgs phase is spontaneously broken SU(2)${}_{global}$, regardless of whether $\langle \phi \rangle$  is zero or non-zero in some gauge. If that idea makes sense, then we must be able to find a gauge-invariant order parameter which is sensitive to the symmetry breaking but insensitive to any gauge choice.  Such an order parameter must be inherently non-local, since we know from the work of Osterwalder and Seiler \cite{Osterwalder:1977pc} that the VEV of local gauge-invariant observables in a gauge-Higgs theory is analytic in the coupling constants, along a path joining the confinement-like to the Higgs regime.  We must also confront the Goldstone theorem:  if a global continuous symmetry is spontaneously broken, how can massless excitations be avoided?  The answer is that
the global symmetry can be broken in the Higgs sector, without actually breaking (and giving rise to Goldstone modes) in the full theory.  

     To explain this point, let us begin by noting that the partition function $Z(\b,\g)$ of the gauge-Higgs theory can be regarded as the weighted sum of partition functions $Z_{spin}(\g,U)$ of a spin system in a background gauge field, i.e.
\beq
            Z(\b,\g) = \int DU ~ Z_{spin}(\g,U) e^{-S_W[U]} \ ,
\label{sumspin}
\eeq
where
\bea
          Z_{spin}(\g,U) &=& \int D \phi ~ e^{-S_H[\phi,U]} \non \\
                                     &=& e^{-\F_H[\g,U]} \ .
\label{spin}
\eea                                     
The only symmetry of the spin system, since $U_\m(x)$ is fixed, is the SU(2)${}_{global}$ symmetry $\phi(x) \ra \phi(x) R$, and
this symmetry may or may not be spontaneously broken, depending on the gauge field configuration $U_\m(x)$.   Our observation is
that the symmetry may be spontaneously broken in every $Z_{spin}(\g,U)$ for which $U$ is  a thermalized configuration, without breaking the symmetry, or introducing a Goldstone mode, in the full theory.  By ``thermalized'' we mean a member of the set of configurations which dominate
the functional integral \rf{sumspin}, samples of which are generated numerically in lattice Monte Carlo simulations.

   How can we tell whether the global symmetry symmetry is spontaneously broken in these spin systems? If we denote the VEV of the $\phi(x)$ field in the background gauge field
as $\overline{\phi}(x;U)$, where
\bea
      \overline{\phi}(x;U)  
      &\equiv& {1\over Z_{spin}(\g,U)} \int D\phi ~ \phi(x) e^{-S_H[\phi,U]}     \ ,
\label{ophi}
\eea
then in general, in a lattice volume $V$,
\beq
           {1\over V} \sum_x ~ \overline{\phi}(x;U) = 0 \ ,
\label{spav}
\eeq
and this is for two reasons.  First, if no gauge is fixed so that $U_\m(x)$ varies wildly in space, then $\phi(x)$ also varies wildly with position, and the spatial average vanishes.  Still, at any given point $x$ it could be that $\overline{\phi}(x;U) \ne 0$.  But this is impossible for the
second reason:  In a finite volume and in the absence of any explicit SU(2)$_{global}$ breaking term, there
can be no spontaneous symmetry breaking, and, since $\overline{\phi}$ transforms under SU(2)$_{global}$ symmetry, it follows that  
$\overline{\phi}(x;U)=0$ at every point.

  But of course real (and therefore finite volume) magnets can be magnetized at low temperatures, and in that case a global symmetry  has been spontaneously broken, despite formal theorems to the contrary.  The signature of a broken symmetry in
a real magnet, in the absence of an explicit source of symmetry breaking such as an external magnetic field, is the existence of 
long-lived metastable states of different but non-zero magnetization, with lifetimes that increase to infinity as $V\ra \infty$.  We can adopt this same principle to study broken symmetry in the spin system defined by \rf{spin}.  The idea is that the Boltmann  probability factor 
$\propto \exp[-S_H[\phi,U]]$ can be generated by long time evolution in a fictitious ``fifth-time'' $t_5$, where the field $\phi(x,t_5)$
evolves according to, e.g., the Langevin equation, or the molecular dynamics approach, or via lattice Monte Carlo simulations.  
In the case of Monte Carlo simulations $t_5$ is discrete, and corresponds to the number of update sweeps through the lattice.  But in any of these methods, the expectation value of an operator $O$ is defined by
\beq
            \overline{O} = \lim_{T_5 \ra \infty} {1\over T_5} \int_0^{T_5} dt_5 ~ O[\phi(x,t_5)] \ .
\eeq
We then use the fifth-time formalism, instead of \rf{ophi}, to define
\beq
             \overline{\phi}(x;U)  = \lim_{T_5 \ra \infty}  \lim_{V\ra \infty} {1\over T_5} \int_0^{T_5} dt_5  ~ \phi(x,t_5) \ ,
\label{t5}
\eeq
with the order of limits as shown.  If $\overline{\phi}(x;U)=0$ at every point, then the symmetry is unbroken, otherwise the 
SU(2)$_{global}$ symmetry is broken spontaneously.  Even if the symmetry is broken, it is still true that the spatial average of
$\overline{\phi}(x;U)$ will vanish in general, as in \rf{spav}.  Moreover, $\overline{\phi}(x;U)$ is gauge-covariant rather than gauge-invariant, transforming as
\beq
             \overline{\phi}(x;g \circ U) = g(x) \overline{\phi}(x;U) \ .
\eeq
However, this quantity has a gauge-invariant modulus 
\beq
|\overline{\phi}(x;U)| =\sqrt{\oh \tr[\overline{\phi}^\dg(x;U)\overline{\phi}(x;U)]} \ ,
\label{mod1}
\eeq
and the spatial average of the modulus is positive if $\overline{\phi}(x;U)$ is non-zero in general.
We therefore define, as our gauge-invariant order parameter, the spatial average
\bea
            \Phi[U] &=& \lim_{T_5 \ra \infty}  \lim_{V\ra \infty} {1\over V} \sum_x \left|{1\over T_5} \int_0^{T_5} dt_5  ~ \phi(x,t_5) \right|  \ ,
\eea
with $\Phi[U]=0$ or $\ne 0$ in the unbroken and spontaneously broken cases respectively.   If at given couplings $\b,\g$  we find that
$\Phi[U] \ne 0$ for gauge field configurations contributing to $Z(\b,\g)$ in the thermodynamic limit, i.e.\ if
\bea
          \langle\Phi \rangle &\equiv& {1\over Z(\b,\g)} \int DU ~ \Phi[U] e^{-(S_W[U]+\F_H[U])} > 0   \ ,                           
\eea 
then by this definition SU(2)$_{global}$ is spontaneously broken in each of the $Z_{spin}$ subsystems, at that point in the $\b-\g$ phase diagram.

    We can now understand the absence of Goldstone modes.  The order parameter for symmetry breaking in a $Z_{spin}(\g,U)$ system
 is the gauge covariant quantity $\overline{\phi}(x;U)$, which vanishes when averaged over gauge-field configurations, i.e.\
 \beq
          \langle  \overline{\phi}(x;U)  \rangle = 0
\eeq
The same can be said of long-range
correlations in various $n$-point functions.  Such long-range correlations only exist, in a theory at fixed $U$ and $\Phi[U] > 0$, in the $n$-point functions of gauge non-invariant operators.  These correlators vanish in the full theory.  To pick a trivial example, the correlator
\beq
             \oh \overline{\tr[\phi^\dg(x) \phi(y)]} = {1\over Z(\g,U)} \int D\phi ~ \oh \tr[\phi^\dg(x) \phi(y)] e^{-S_H[\phi,U]}
\eeq
may have long range correlations for a particular gauge field $U$ with $\Phi[U] > 0$, but this quantity vanishes when integrating over all gauge fields,
\beq
\langle \overline{\tr[\phi^\dg(x) \phi(y)]} \rangle = 0 \ ,
\eeq
as does $\langle \tr[\phi^\dg(x) \phi(y)] \rangle$.  
One could, of course, construct a gauge-invariant quantity such as
\beq
G(x,y) =\langle \overline{\tr[\phi^\dg(x) U(x,y)\phi(y)]} \rangle \ ,
\eeq
where $U(x,y)$ is a Wilson line with endpoints $x,y$,
but there is no particular reason why this quantity should have a power-law falloff.  The point here is that long-range correlations in
the individual $Z_{spin}(\g,U)$, which are due
to the Goldstone theorem, must cancel out in the full theory.  

   But the absence of Goldstone modes does not mean that gauge-Higgs theory in the ``broken'' phase (meaning that all the non-negligible
spin systems are in the broken phase), is qualitatively similar to gauge-Higgs theory in the unbroken phase.  We will elaborate on how these
phases can differ in section V.


\subsection{SU(2)$_{global}$ in unitary gauge}

   One might wonder what happens to the SU(2)$_{global}$ symmetry in unitary gauge, where there is no longer any freedom to transform
$\phi$.  In fact nothing happens; the symmetry is still there.  Let us fix to $\phi=\mathbbm{1}$.  Then
\beq
          Z = \int DU \exp[-S_W + \g \sum_{x,\m} \oh \tr U_\m(x)] \ .
\eeq
Now let $F[U]=0$ be any gauge-fixing condition, and we insert unity in the usual way:
\bea
         Z &=& \int DU \left\{ \D_{FP}[U] \int Dg \d(F[g \circ U]) \right\} \non \\
             & & \qquad \times \exp[-S_W + \g \sum_{x,\m} \oh \tr U_\m(x)]  \non \\
            &=& \int DU  \D_{FP}[U]  \d(F[U]) e^{-S_W}  \non \\
            & & \qquad \times \int Dg \exp[ \g \sum_{x,\m} \oh \tr [g^\dg(x) U_\m(x) g(x+\hat{\m})] \non \\
            &=& \int DU ~ \D_{FP}[U]  \d(F[U]) Z_{spin}(\g,U) e^{-S_W} \ .
\eea
The last line is eq.\ \rf{sumspin} in the gauge $F[U]=0$.  Since the order parameter $\Phi$ for symmetry-breaking in
$Z_{spin}(\g,U)$ is gauge-invariant, we recover the original formulation, with $\phi(x)$ replaced by $g(x)$.    

\subsection{Numerical procedure}
  
   We calculate $\langle \Phi \rangle$ by  a Monte Carlo-within-a-Monte Carlo procedure.  That is to say, the usual update sweeps involve
sweeping site by site through the lattice, and updating the four link variables and the Higgs field at each site.  Since both the link and scalar field variables are elements of the SU(2) group, the updates of both types of variables can be carried out using the Creutz heat bath method.  In this method one seeks to stochastically generate SU(2) elements $G$ according to a probability distribution
\beq
           dP(G) \propto e^{\oh \tr[GA]} dG \ ,
\label{hbath}
\eeq 
where $A$ is a fixed matrix proportional to an SU(2) group element.  For updating a link variable $G=U_\m(x)$, we have
\bea
           A &=& \b \sum_{\n \ne \m} \Bigl\{U_\n(x+\hat{\m}) U^\dg_\m(x+\hat{\n})U^\dg_\n(x)  \non \\
               & &                              + U^\dg_\n(x+\hat{\m}-\hat{\n}) U^\dg_\m(x-\hat{\n})U_\n(x-\hat{\n}) \Bigr\} \non \\
               & & + \g \phi(x+\hat{\m}) \phi^\dg(x)  \ ,
\eea
while for updating a scalar field variable $G=\phi(x)$ we use
\beq
          A = \g \sum_\m \Bigl( \phi^\dg(x-\hat{\m}) U_\m(x-\hat{\m}) + \phi^\dg(x+\hat{\m}) U^\dg_\m(x) \Bigr)
\eeq
The heat bath procedure for generating group elements $G$ in a probability distribution \rf{hbath} is described in standard texts such
as \cite{Gattringer:2010zz}, and in the seminal paper by Creutz \cite{Creutz:1980zw}.    

The data-taking sweep, however,  is a simulation of the spin system \rf{spin}, and entails $n_{sw}$ sweeps through
the lattice, updating only the Higgs field by the heat bath method, while keeping the gauge field fixed.  In the course of this data-taking sweep, on a
finite lattice volume $V$, we measure
\beq
           \Phi_{n_{sw},V}[U] = {1\over V} \sum_x \left|{1\over n_{sw}} \sum_{t_5=1}^{n_{sw}} \phi(x,t_5) \right| \ ,
\label{bigphi}
\eeq
where $\phi(x,t_5)$ is the Higgs field at point $x$ after $t_5$ update sweeps, holding the $U$ field fixed.  The quantity we 
would like to estimate is the limiting value
\beq
      \langle \Phi\rangle = \lim_{n_{sw}\ra \infty} \lim_{V\ra \infty} \langle \Phi_{n_{sw},V}[U] \rangle \ ,
\eeq
again with the order of limits as shown.  In the infinite volume limit we expect, on general statistical grounds, that
\beq
          \langle \Phi_{n_{sw},\infty}[U] \rangle \approx \langle \Phi \rangle + {\mbox{const.} \over \sqrt{n_{sw}}} \ .
\label{extrapolate}
\eeq
In the unbroken phase, with $\langle \Phi \rangle=0$, this behavior would also hold at finite volume.  In the broken phase,
however, we expect $\langle \Phi_{n_{sw},V}[U] \rangle \approx \langle \Phi_{n_{sw},\infty}[U] \rangle$ to only hold for $n_{sw}$ smaller than
the lifetime $T_{meta}(V)$ of the metastable state, and then to go to zero as $n_{sw}$ increases beyond $T_{meta}(V)$.
So on a finite volume we must use \rf{extrapolate} to extrapolate, from a set of values $\{\langle \Phi_{n_{sw},V}[U] \rangle\}$ computed
at $n_{sw} < T_{meta}(V)$ to the $n_{sw} \ra \infty$ limit,
checking that $T_{meta}(V)$, where the linear extrapolation breaks down, increases with lattice volume $V$, and that the extrapolated
estimate for $\langle \Phi \rangle$ converges as $V$ increases.

   To pin down the point of transition, it is also helpful to introduce a gauge-invariant quantity which functions as a susceptibility:
\beq
             \chi =  V \left\langle \Bigr|{1\over V} \sum_x  \tr[\vph(x;U)\{\phi(x) - \overline{\phi}(x;U)\}] \Bigr|^2 \right\rangle \ ,
\label{chi}
\eeq
where we have defined a gauge covariant, unimodular field
\beq
\vph(x;U) = {\overline{\phi}(x;U) \over |\overline{\phi}(x;U)|}  \ .
\eeq
The transition point, at fixed $\b$, is identified with the value of $\g$ where $\chi$ is maximized.

   In the unbroken phase $\vph(x,U)$ is $0/0$, strictly speaking, and $\chi$ has to be defined again in a fifth-time formalism.  Let
$\phi(x,t_5)$ denote the Higgs field configuration obtained after $t_5$ update sweeps in the spin system simulation at fixed $U$ and
lattice volume $V$. Then define
\bea
    \overline{\phi}_{n_{sw},V}(x;U) &=& {1\over n_{sw}} \sum_{t_5=1}^{n_{sw}} \phi(x,t_5)  \non \\
    \vph_{n_{sw},V}(x;U) &=& {\overline{\phi}_{n_{sw},V}(x;U) \over |\overline{\phi}_{n_{sw},V}(x;U)|} \ ,
\label{phis}
\eea
and construct, in terms of these quantities,
\bea
\chi &=&
  V \left\langle  \Bigl| {1\over V} \sum_x {1\over n_s} \sum_{t_5=n_{sw}+1}^{n_{sw}+n_s} 
      \tr[\vph_{n_{sw},V}(x;U) \right.  \non \\
     & & \left.  \qquad \qquad \{\phi(x,t_5)- \overline{\phi}_{n_{sw},V}(x;U)\}] \Bigr|^2 \right\rangle \ .
\label{csus}
\eea
Apart from the finite lattice volume $V$, this definition involves a choice of $n_{sw}$ for defining a gauge covariant
field $\vph_{n_{sw},V}(x;U)$, and a choice of $n_s \ll n_{sw}$ for the estimate of susceptability.  The parameter $n_{sw}$ is chosen to
be large enough to avoid substantial statistical errors, but small enough so that, in the broken phase, we do not have 
$\overline{\phi}_{n_{sw},V}(x;U)$ much smaller than the limit in \rf{t5}  just due to the formal absence of symmetry breaking 
in a finite volume.   Likewise, the choice of $n_s$ balances  the requirement of small statistical errors ($n_s$ large), with a condition
that $\overline{\phi}_{n_{sw},V}(x;U)$ and $\overline{\phi}_{n_{sw}+n_s,V}(x;U)$ do not differ appreciably.
In practice we have used $n_{sw}=900$ and $n_s=100$ in computing $\chi$. 

   We have found that $\chi$ defined in this way is very useful in practice for locating the transition point, but we do not have a
rigorous argument for why this works so well.  The proper definition of the transition point is that $\langle \Phi \rangle$ is zero, in the appropriate limits, below the transition point, and is non-zero above that point.  We have found that this condition is satisfied by the
transition point suggested by the peak in $\chi$, in every case we have examined.

\subsection{Landau Gauge}
 
   We must check whether the gauge-invariant symmetry breaking criterion $\Phi[U] > 0$ is a Landau gauge criterion \cite{ Kennedy:1986ut,Langfeld:2002ic,Caudy:2007sf} in disguise.

    When $\Phi[U] > 0$ in the appropriate limits, it means that the Higgs field fluctuates preferentially around one of a set of
field configurations, related by SU(2)$_{global}$ transformations, in the infinite volume limit.  It is natural to suppose that,
in a given background gauge field, $\phi(x)$ fluctuates around the configuration which minimizes the Higgs action $S_H$, and
which therefore maximizes
\beq
         \sum_{x,\m} \oh \mbox{Tr}[\phi^\dg(x) U_\m(x) \phi(x+\widehat{\m})]  \ .
\eeq
That is, after all, the starting assumption of any perturbative expansion.

    Let $\phi_{max}(x)$ be this maximizing configuration.  Then $g(x)=\phi_{max}(x)$ is a gauge transformation which takes $U_\m(x)$ into
Landau gauge.  It follows that if $U_\m(x)$ is already in Landau gauge, then $g(x)=\phi_{max}(x)$ is the gauge transformation which
preserves the Landau gauge condition, and this is the remnant symmetry of Landau gauge, namely the  
transformations $g(x)=g$ which are independent of position.  So we might expect, in the broken phase of the spin system \rf{spin}
with $U$ in Landau gauge, that $\phi(x)$ fluctuates around one of the maximizing configurations $\phi(x)=g \in $ SU(2).   

    The order parameter proposed in \cite{Caudy:2007sf,Langfeld:2002ic} was devised to detect the breaking of remnant symmetry in Landau gauge.  We define, in lattice volume $V$
\bea
      \Omega_V[U] &=&  \left| {1\over V} \sum_x   ~ \phi_L(x) \right|^2  \ ,
\label{Landau}
\eea
where the subscript $L$ in $\phi_L(x)$ indicates that $\phi(x)$ is computed Landau gauge.  
Note that in \rf{Landau} the modulus is taken after the spatial average, whereas in the definition of $\Phi[U]$ the modulus is
taken prior to the sum over position.  We can pin down the transition point from the peak in susceptability
\beq
\chi_L = V ( \langle  \Omega^2_V \rangle - \langle \Omega_V \rangle^2) \ .
\label{chiL}
\eeq

    Since $\Phi[U]$ is gauge-invariant, it can always be evaluated in Landau gauge, and if $\Phi[U] >0$, it means that
$\phi(x)$ fluctuates around some fixed configuration.  One would imagine that this configuration would be a fixed
group element, constant in spacetime, in which case $\Omega[U]$ is also non-zero, and there is then no real difference
between the two criteria.  The flaw in the argument is that there exist many Gribov copies in Landau gauge, and if $U$ is fixed
to one of them, then there exists a gauge transformation $g'(x)$ to some other copy, and therefore $\phi_{max}(x)=g'(x)$ is also
a local maximum of $S_H$.  It may be that $\phi(x,t_5)$ fluctuates around a $\phi_{max}(x)$ of this kind, whose spatial average
vanishes.  In that case it is possible that both $\langle \Phi \rangle > 0$ and $\langle \Omega \rangle = 0$ hold simultaneously
for some range of couplings, a possibility which we now show is confirmed by the data.
 
\begin{figure*}[t!]
\subfigure[~gauge invariant susceptability]  
{   
 \label{fig1a}
 \includegraphics[scale=0.65]{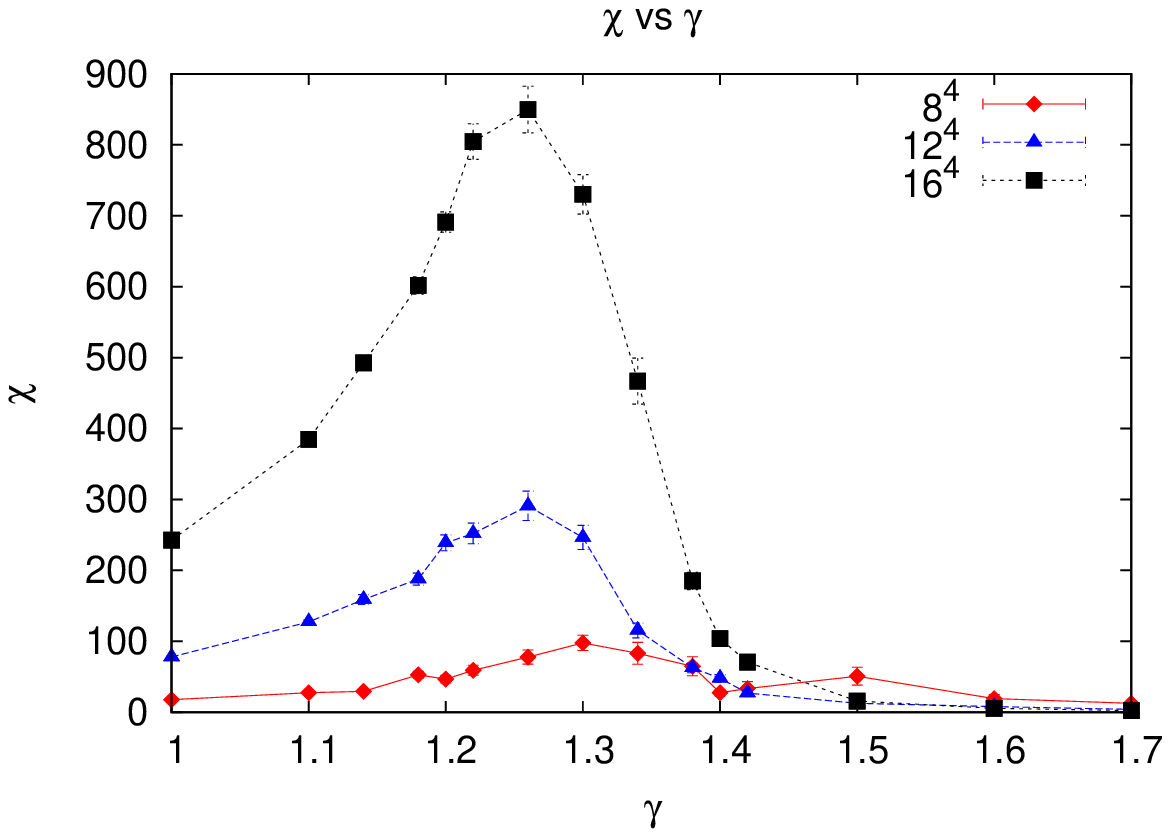}
}
\subfigure[~Landau gauge susceptability]  
{   
 \label{fig1b}
 \includegraphics[scale=0.65]{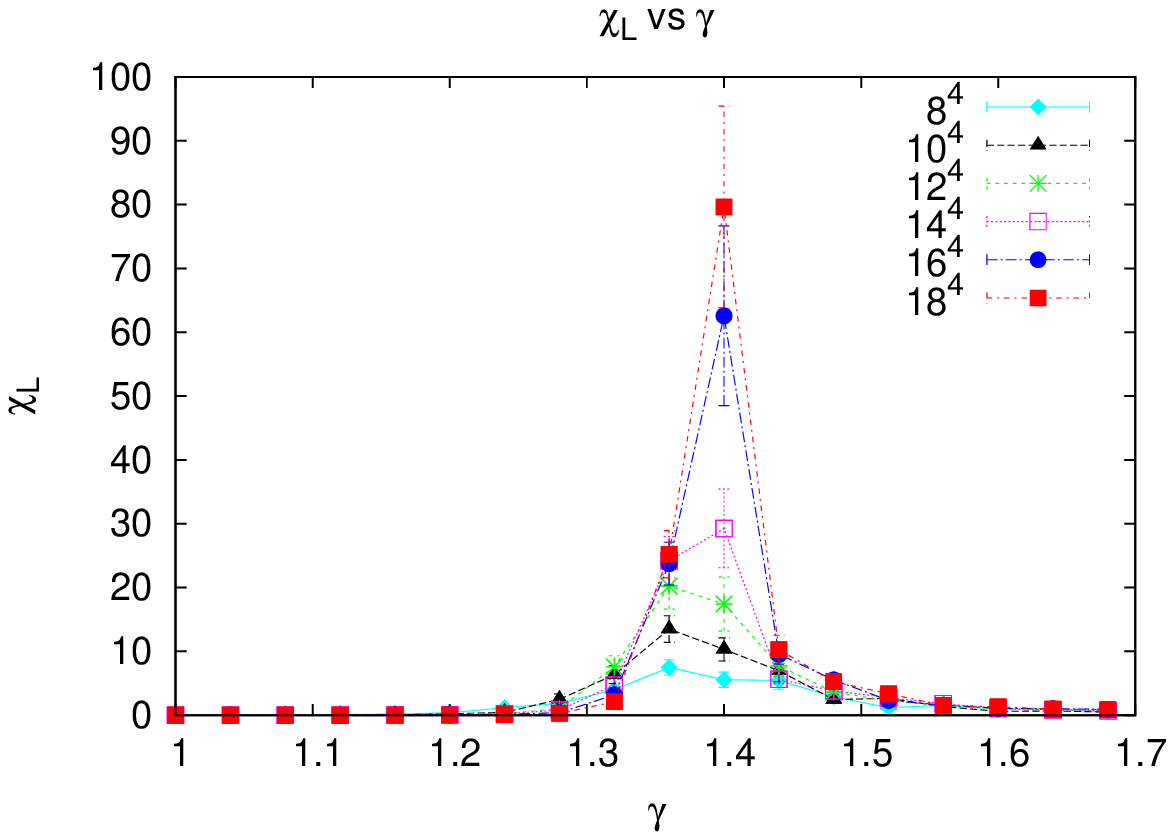}
}
\caption{(a) Gauge-invariant susceptibility $\chi$ vs.\ $\g$, and (b) Landau gauge susceptibility $\chi_L$ vs.\ $\g$, both at
$\b=1.2$ and various lattice volumes.  Note that the peaks in these two susceptibilities occur at different places, i.e. at $\g=1.28$ for the gauge-invariant transition, and at $\g=1.4$ for the Landau gauge transition.} 
\label{suscept12}
\end{figure*}
   
\begin{figure*}[htb]
\subfigure[~below the transition, $\g=1.2$]  
{   
 \label{fig2a}
 \includegraphics[scale=0.65]{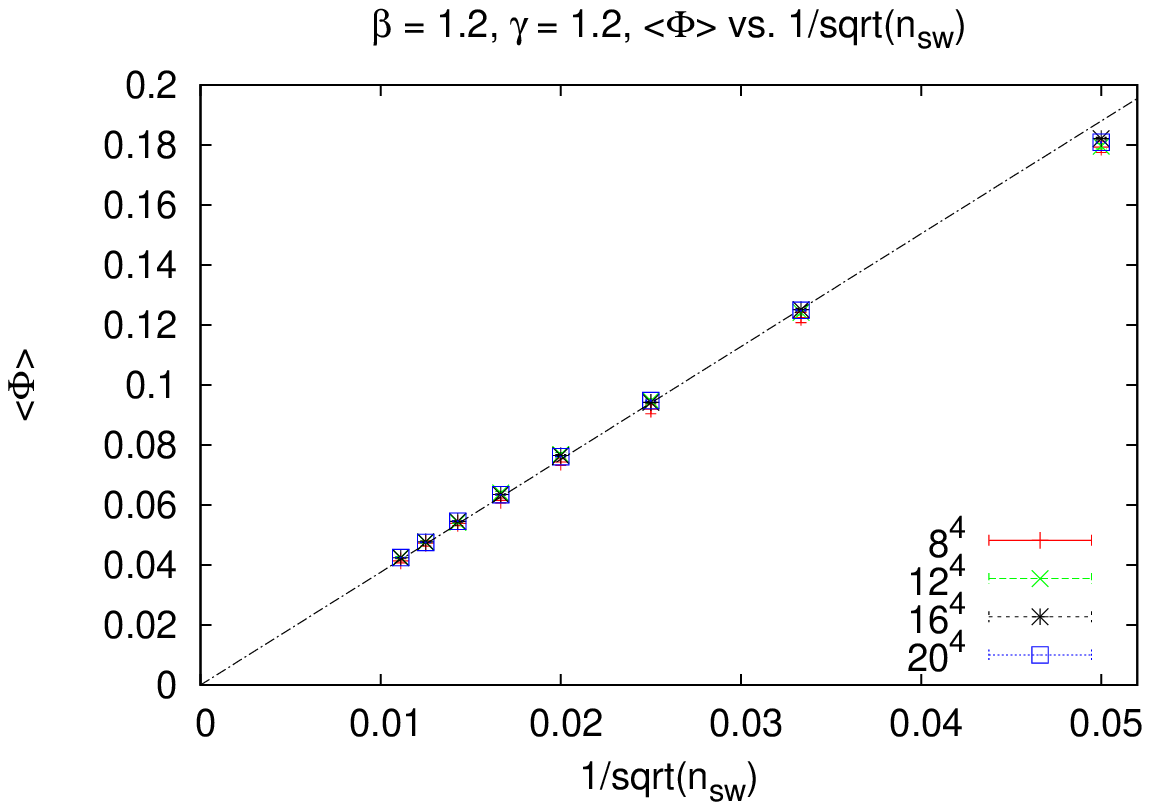}
}
\subfigure[~above the transition, $\g=1.35$]  
{   
 \label{fig2b}
 \includegraphics[scale=0.65]{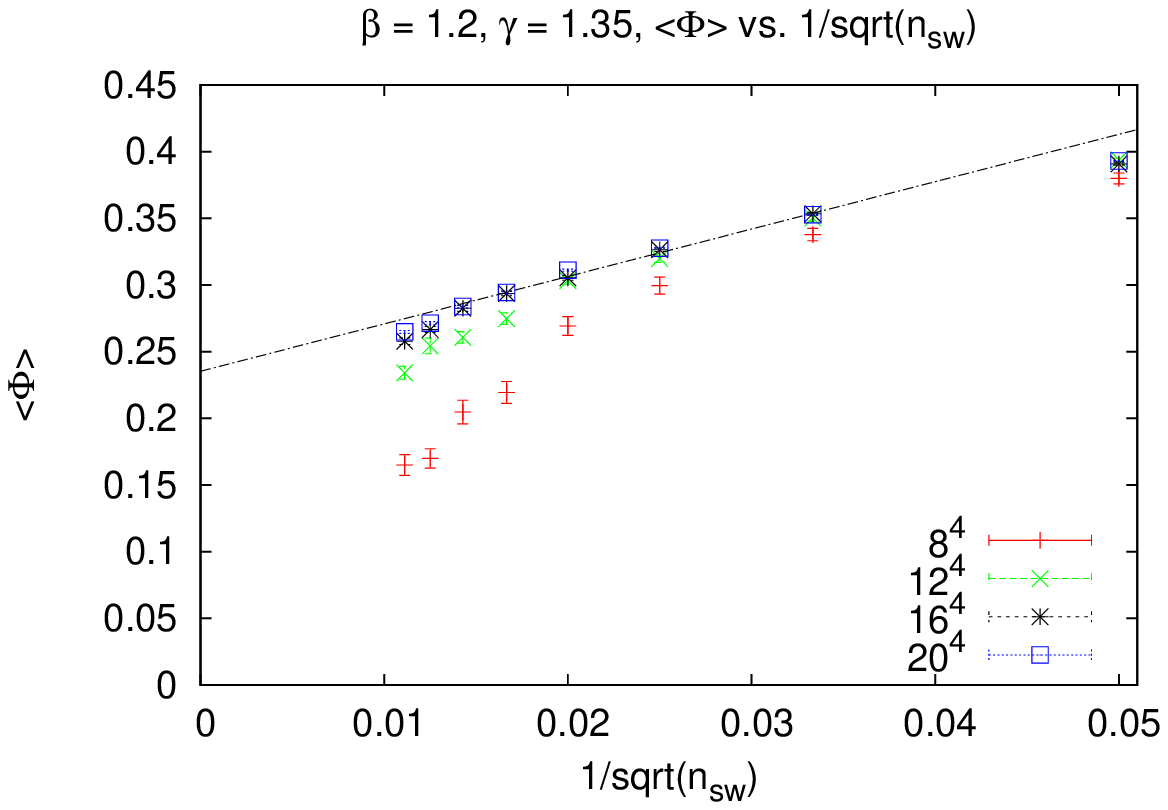}
}
\caption{Gauge invariant order parameter $\Phi$ vs.\ $1/\sqrt{n_{sw}}$, where
$n_{sw}$ are the number of sweeps carried out on the matter field at fixed gauge field.
The data is for $\b=1.2$ at lattice volumes $8^4,12^4,16^4,20^4$.  (a) below
the transition, at $\g=1.2$; (b) above the transition, at $\g=1.35$. Note the convergence, in subfigure (b),
to a straight line with non-zero intercept on the $y$-axis, as lattice volume increases.}
\label{fig2}
\end{figure*}   
   
\begin{figure}[h!]
 \includegraphics[scale=0.7]{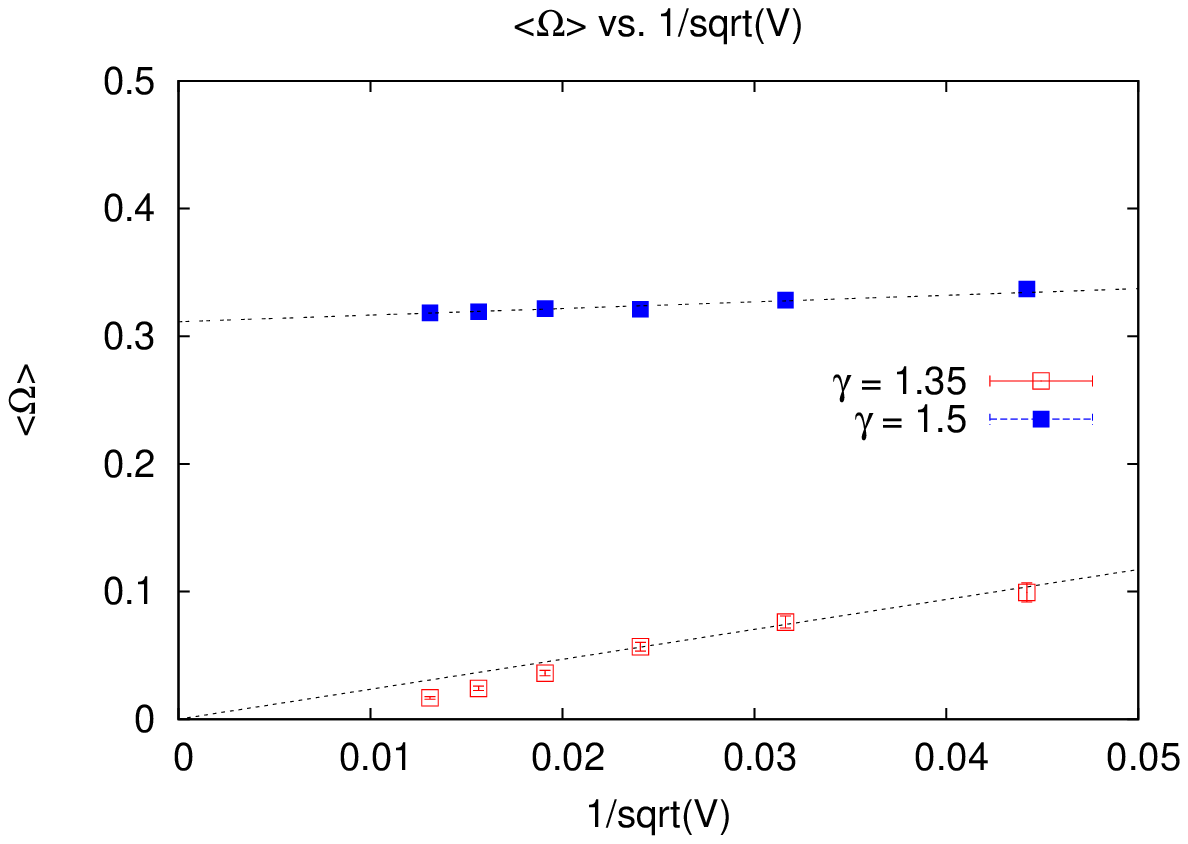}
\caption{Landau gauge order parameter $\Omega$ vs.\ $1/\sqrt{V}$, where $V$ is the lattice volume,
at $\b=1.2$.  Data is shown below the transition, at $\g=1.35$, and above the transition, at $\g=1.5$. } 
\label{fig3}
\end{figure} 
   
\section{Numerical Results}

     There have been many numerical studies of the phase structure of the SU(2) gauge-Higgs model described by \rf{Sgh}, and these
have found a transition line in the $\b-\g$ plane terminating at a finite $\b \approx 2$.  In early studies 
\cite{Lang:1981qg,*Campos:1997dc,*Langguth:1985dr,*Jersak:1985nf} this was considered to be a line of first order transition, but according to the most recent work \cite{Bonati:2009pf} it is only a region of sharp
crossover behavior, up to at least $\b=2.725$.  It is possible that true first order transitions appear at $\b>2.725$. This is all in accordance with the Osterweiler-Seiler theorem \cite{Osterwalder:1977pc}.  No ``Coulomb'' region, corresponding to a $1/R$ potential between static sources, has been found in the phase diagram, although in principle such a region is not ruled out a priori \cite{Fradkin:1978dv}.  String breaking in the confinement-like region of SU(2) gauge-Higgs models has been reported in \cite{Knechtli:1998gf,*Philipsen:1998de}.

   Our procedure is as follows:  After thermalization (up to 4000 updating sweeps on a $20^4$ lattice), we take data after every 100 updating sweeps.  In each data-taking sweep
we begin by saving the lattice configuration, fixing to Landau gauge, and computing $\Omega(U)$ in \rf{Landau}, which is used to
compute the Landau gauge susceptability $\chi_L$ \rf{chiL}.  The lattice is then restored to the saved configuration.  This is followed by
a Monte Carlo within a Monte Carlo; meaning that we hold the gauge link variables fixed, and update only the Higgs field from $t_5=1$ to $t_5=n_{sw}$ sweeps.  Denote the Higgs field at point $x$ and the Higgs-only update sweep $t_5$ as $\phi(x,t_5)$.  We compute 
$\Phi_{n_{sw},V},\overline{\phi}_{n_{sw},V}(x;U),\vph_{n_{sw},V}(x;U)$ according to eqs.\ \rf{bigphi} and \rf{phis} respectively. Our simulations
were carried out on volumes $8^4,12^4,16^4,20^4$, at each of $n_{sw}= 100 N^2, ~ N=2,3,...,12$.  We then carried out the Higgs-only
updates for a further $n_s=100$ sweeps, to calculate $\chi$ in eq.\ \rf{csus}.  Finally, the lattice is restored to the saved configuration.  For the
largest $20^4$ lattice we collected 80 data sets.\footnote{Error bars were computed from a simple standard deviation of the mean; we did not check autocorrelations in this study.} 

     We begin with a display of the susceptibilities $\chi, \chi_L$ vs.\ $\g$ at $\b=1.2$ in Fig.\ \ref{suscept12}.  It is known, from 
\cite{Lang:1981qg,*Campos:1997dc,*Langguth:1985dr,*Jersak:1985nf} and from \cite{Caudy:2007sf},
that there is no thermodynamic transition in $\g$, or even a sharp crossover, at this fixed value of $\b$.
At $\b=1.2$ there seems, however, to be a gauge-invariant symmetry breaking transition at $\g=1.28$, and the Landau transition is
at $\g=1.4$.  In Fig.\ \ref{fig2} we plot the corresponding order parameter $\langle \Phi \rangle$ vs.\ $1/\sqrt{n_{sw}}$, at various
lattice volumes, below (Fig.\ \ref{fig2a}) and above (Fig.\ \ref{fig2b}) the transition point, at $\g=1.2$ and $\g=1.35$ respectively, and we see that the order parameter behaves as expected, falling to zero as $n_{sw} \ra \infty$ below the transition.  Above the transition the data indicates that
$\langle \Phi \rangle \ra 0$ in this limit {\it at fixed volume}, but it can also be seen that the onset
of the drop towards zero increases with lattice volume, consistent with $\langle \Phi \rangle > 0$ in the appropriate pair of
$V \ra \infty, n_{sw} \ra \infty$ limits.  Likewise, the order parameter $\langle \Omega \rangle$ shown in Fig.\ \ref{fig3} 
for the Landau transition just below ($\g=1.35$) and just above ($\g=1.5$) the transition behave as expected, falling to zero with $1/\sqrt{V}$ below the transition, and converging to a non-zero constant at large $V$ above the transition.  The point to notice here is that
at $\b=1.2, \g=1.35$ we have exactly the situation noted last section, i.e.\ there is
a region in the phase diagram where $\langle \Phi \rangle > 0$ and $\langle \Omega \rangle = 0$.  From this type of data we conclude 
that the gauge invariant criterion for SU(2)$_{global}$ symmetry breaking is not the same as the Landau gauge criterion.

\begin{figure}[htb]
 \includegraphics[scale=0.7]{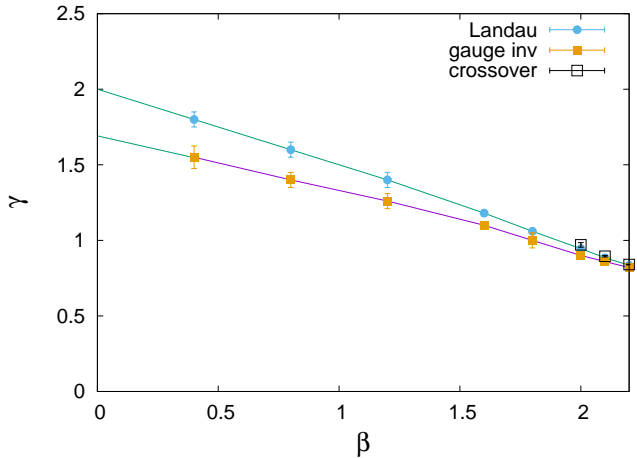}
\caption{Transition line (square points) for the gauge-invariant global SU(2) symmetry described in the text.  The transition line
for remnant gauge symmetry breaking in Landau gauge (circles) is shown for comparison, along with points at $\b \ge 2.0$ (open squares)
 where we find a sharp thermodynamic crossover.} 
\label{phase}
\end{figure}

    From the peaks in $\chi$ and $\chi_L$ we can locate the transition lines for the gauge-invariant symmetry breaking
transition, and for the Landau gauge transition, and these are shown in Fig.\ \ref{phase}.  The Landau gauge transition line was
previously found in \cite{Caudy:2007sf}, and our present result for that line agrees with the older calculation.  The
gauge-invariant symmetry breaking transition line is a new result.   

    At the larger $\b \ge 2.0$ values, where there is a relatively sharp thermodynamic crossover in $\g$, we can find the crossover point from a peak in the plaquette susceptability.  Let
\bea
       E_P &=&   {1\over 6 L^4} \sum_{plaq} \oh \mbox{Tr}[U_\m(x)U_\n(x+\hat{\m})U_\m^\dg(x+\hat{\n}) U^\dg_\n(x)]  \non \\
       E_H &=&    {1\over 4 L^4}  \sum_{x,\m} \oh \mbox{Tr}[\phi^\dg(x) U_\m(x) \phi(x+\widehat{\m})] \ ,
\eea
be the plaquette energy density and average Higgs energy density, respectively.  Then the plaquette susceptability is
\beq
       \chi_P = {\pa \langle E_P \rangle \over \pa \gamma} = 4 L^4 (\langle E_P E_H \rangle - \langle E_P \rangle \langle E_H \rangle)
\eeq
The location of the peaks in this susceptability, which lie on the thermodynamic crossover line (at $\b \ge 2.0$) originally found in 
\cite{Lang:1981qg}, are also displayed in Fig.\ \ref{phase}.
  
\section{U(1) symmetry breaking in SU(3) gauge-Higgs theory}

    The SU(2)$_{global}$ symmetry in SU(2) gauge-Higgs theory is in some sense accidental and there is, in the general case, no 
 SU($N$)$_{global}$ symmetry in an SU($N$) gauge-Higgs theory. This is simply because the mapping of a Higgs multiplet to a group element, as in \rf{map}, does not generalize to SU($N$) theories.  There does exist, however, a global U(1) symmetry in SU($N>2$) 
gauge-Higgs theories, with 
\beq
   S_H[U,\phi] = - \gamma \sum_{x,\m} \mbox{Re}[\phi^\dg(x) U_\m(x) \phi(x+\widehat{\m})] \ ,
\label{Shiggs2}
\eeq
and where the unimodular Higgs field transforms in the fundamental representation of SU($N$).  This action is invariant, as pointed
out by Maas et al.\  \cite{Maas:2016ngo,*Maas:2017xzh,*Maas:2017pcw}, under
the U(1) transformations
\beq
              \phi(x) \ra e^{i\th} \phi(x)  \ ,
\label{U1}
\eeq
and our point is that this global symmetry, like any global symmetry, can be spontaneously broken. 

    The order parameter for the spontaneous symmetry breaking of the global symmetry \rf{U1} in the spin system \rf{spin} is 
essentially identical to the $\langle \Phi \rangle$ order parameter defined in section \ref{sec2}, changing only the definition
of the gauge invariant modulus
\beq
|\overline{\phi}(x;U)| =\sqrt{\overline{\phi}^\dg(x;U)\overline{\phi}(x;U)} \ ,
\label{mod2}
\eeq    
where a dot product of color indices, rather than a trace, is implied.  As before, $\langle \Phi \rangle = 0$ means that the
global symmetry is unbroken, while $\langle \Phi \rangle > 0$ implies spontaneous breaking of the
global U(1) symmetry.

\begin{figure}[htb]
 \includegraphics[scale=0.7]{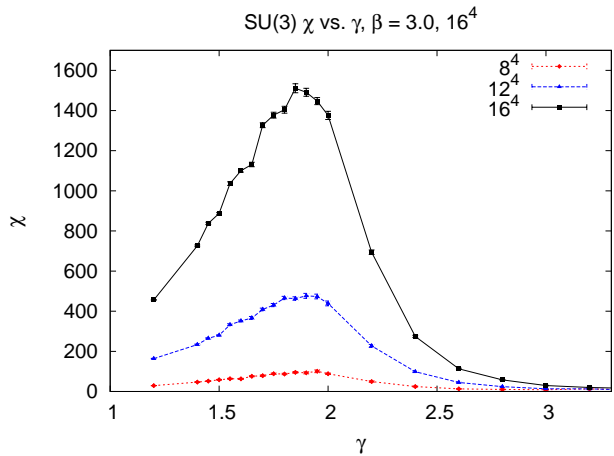}
\caption{Gauge-invariant susceptability $\chi$ vs.\ $\g$ for SU(3) gauge-Higgs theory at $\b=3.0$, and lattice volumes $8^4,12^4,16^4$.}   
\label{b3}
\end{figure}

\begin{figure}[htb]
 \includegraphics[scale=0.7]{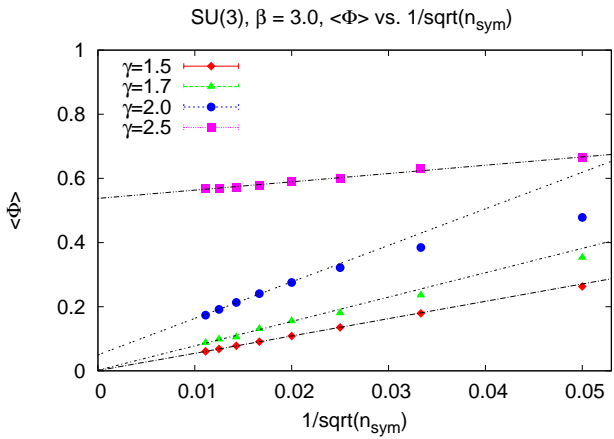}
\caption{Gauge invariant SU(3) order parameter $\Phi$ vs.\ $1/\sqrt{n_{sw}}$ at $\b=3.0$ on a $16^4$ lattice volume.  Below the transition at $\g=1.85$, the data extrapolates to zero as $n_{sw}\ra \infty$.  Above the
transition, the data extrapolates to non-zero values.} 
\label{opfig}
\end{figure} 

\begin{figure}[htb]
 \includegraphics[scale=0.7]{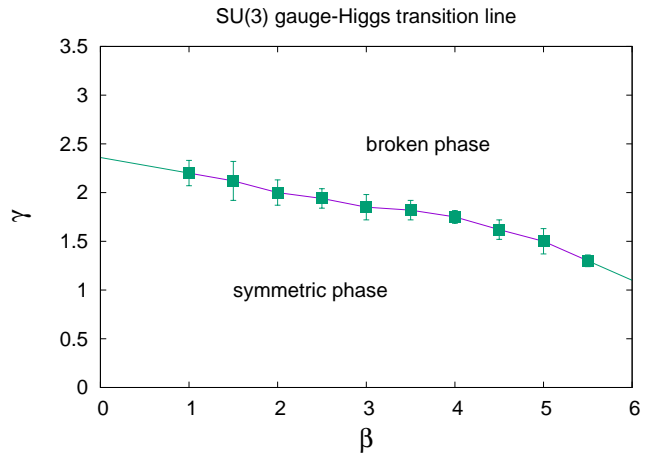}
\caption{Gauge-invariant transition line for global U(1) symmetry breaking in SU(3) gauge-Higgs theory.} 
\label{su3gb}
\end{figure} 

    We have computed the transition line in the $\b-\g$ coupling plane for SU(3) gauge-Higgs theory, with the action consisting
of the Wilson action
\beq
      S_W = -  \beta \sum_{plaq} {1\over 3} \mbox{ReTr}[U_\m(x)U_\n(x+\hat{\m})U_\m^\dg(x+\hat{\n}) U^\dg_\n(x)] 
\eeq
plus $S_H$ in \rf{Shiggs2}, and a unimodular Higgs field.  The numerical ``Monte Carlo within a Monte Carlo'' procedure is essentially
the same as the one described in the previous section for the SU(2) case.   The only difference is that updates of link and scalar field
variables are carried out via the Metropolis algorithm.  The first step is to generate an SU(3) matrix close to the identity element, and
this is done by generating stochastically three SU(2) matrices, which are each embedded in a $3\times 3$ matrix, as described in section 4.2.3 of ref.\ \cite{Gattringer:2010zz}.  Each SU(2) submatrix is generated by the heat bath method, with $A=\a \mathbbm{1}$ in \rf{hbath}.  The product of these three matrices is an SU(3) matrix $G$.  When updating a link variable we generate a trial link variable $U_{try} = G U_\m(x)$, and then compute the change in the action $\D S = \D S_W + \D S_H$ generated by the trial link, which is
then accepted or rejected according to the Metropolis algorithm.  Similarly, when updating the scalar field, which is a unimodular 3-vector, we generate a trial variable $\phi_{try} = G \phi(x)$, compute the change in the Higgs action $\D S_H$, and accept or reject by the Metropolis rule.  We adjust the spread of the (stochastically generated) SU(2) matrices away from the identity matrix by adjusting the parameter $\a$.  This parameter is assigned different values for the link and scalar field updates, in order to obtain an acceptance rate in the Metropolis algorithm of about 50\%.

The transition points are located by computing the susceptibility $\chi$ of \rf{chi} at fixed $\b$ over a range of $\g$ and lattice volumes, and identifying the transition point as the location of the peak, as shown in Fig.\ \ref{b3} at $\b=3.0$.  In this case the transition is at $\b \approx 1.85$.  We also check that $\langle \Phi \rangle \ra 0$ as $n_{sw} \ra \infty$ below the
the transition, while $\langle \Phi \rangle$ extrapolates to a non-zero value above the transition.
This is illustrated at $\b=3.0$ in Fig.\ \ref{opfig}.  The transition line in the $\b-\g$ coupling plane, for $0<\b<5.6$ is shown in Fig.\ \ref{su3gb}.

    In the case of compact U(1) gauge-Higgs theory, with a single-charged scalar field, the additional symmetry is also global U(1), and
it is that symmetry which can be spontaneously broken.  This point seems to have been made previously, in connection with 
superconductivity, by Greiter \cite{GREITER2005217}.  It would be interesting to explore the phase diagram of compact scalar QED with a singly charged matter field, which contains a massless as well as confinement-like and Higgs regions, in connection with the broken vs.\
unbroken realization of the global U(1) symmetry.  We leave this for a future investigation. 

\section{Separation-of-charge confinement}

   Given that there exists a gauge-invariant global symmetry in gauge-Higgs theory which is realized in either a broken
or unbroken phase, the obvious question is what physical property distinguishes these two phases.  Our conjecture is that
the symmetry-breaking transition corresponds to a transition between two different types of confinement, which in a previous
article \cite{Greensite:2017ajx} we have referred to as C- and S$_c$-confinement.

  In any SU($N$) gauge theory with matter fields in the fundamental representation, such as  QCD or gauge-Higgs theories, Wilson loops have a perimeter-law falloff asymptotically, and Polyakov lines have a non-zero vacuum expectation value. So what does it mean to say that such theories (QCD in particular) are confining?  Historically, confinement was taken to mean the absence of free quarks in the QCD spectrum, and more generally confinement is defined as the color neutrality of the asymptotic particle spectrum.  We will refer to this property as ``C-confinement.''  As we have already noted in the Introduction, gauge-Higgs theories in the Higgs regime, where there are
no linearly rising Regge trajectories, no string formation and no string breaking, and only Yukawa forces, are also confining by this
definition.  

   On the other hand, in a pure SU($N$) gauge theory, there is a different and stronger meaning that can be assigned to the word ``confinement," which goes beyond C-confinement.  Of course the spectrum of pure gauge theories consists of only color 
 neutral objects, i.e.\ glueballs.  But such theories also have the property that the static quark potential rises linearly or, equivalently, that large planar Wilson loops have an area-law falloff.  It is reasonable to ask if there is any way to generalize this property to gauge theories with matter in the fundamental representation, and such a generalization was proposed in ref.\ \cite{Greensite:2017ajx}.  It begins by noting that
the Wilson area-law criterion in a pure gauge theory is equivalent to the following statement, which we call ``separation of charge" confinement
or ``S$_c$" confinement.  We consider a class of functionals $V(\vx,\vy;A)$ of the gauge field which transform bi-covariantly under
a gauge transformation $g(x)$, i.e.
\beq
          V^{ab}(\vx,\vy;A) \ra  g^{ac}(\vx,t) V^{cd}(\vx,\vy;A) g^{\dg db}(\vy,t) \ ,
\label{bicovariant}
\eeq 
and then define \\

\noindent\fbox{
    \parbox{8cm} {
\ni \underline{\bf S$_c$-confinement} \\

Let $E_V(R)$, with $R=|\vx-\vy|$ be the energy (above the vacuum energy) of a state
\beq
        \Psi_V \equiv  \overline{q}^{a+}(\vx) V^{ab}(\vx,\vy;A) q^{b+}(\vy) \Psi_0 \ ,
\label{Vstate}
\eeq
where $q^+,\overline{q}^+$ represent creation operators for static quark/antiquark color charges, and
$\Psi_0$ is the vacuum state.  \textbf{\textit{S${}_c$-confinement}} means that there exists an asymptotically linear function $E_0(R)$, i.e.\
\beq
    \lim_{R\ra \infty} {dE_0 \over dR} = \s >0 \ ,
\eeq
such that
\beq
        E_V(R) \ge E_0(R) 
\label{criterion}
\eeq
for any choice of bi-covariant $V(\vx,\vy;A)$. 
}}

\bigskip

\ni In a pure gauge theory, S$_c$-confinement is equivalent to the Wilson area law criterion, with $E_0(R)$ the static quark potential and $\s$ the asymptotic string tension.  \\

   Our proposal in \cite{Greensite:2017ajx} is that S$_c$-confinement should also be regarded as the confinement criterion in gauge+matter theories. The crucial element is that the bi-covariant operators $V^{ab}(\vx,\vy;A)$ {\it must depend only on the gauge field A} at a fixed time, and not on the matter fields.  Excluding matter fields from $V^{ab}(\vx,\vy;A)$ means that we are dealing with a subclass of physical states $\Psi_V$ which really correspond to two separated color charges, rather than two separated color neutral objects.  The question that is addressed by this exclusion is whether (i) a non-confining static quark potential is due exclusively to string-breaking effects by matter fields, or whether instead (ii) a non-confining state can be constructed without any appeal to string breaking.  Case (i) is S$_c$-confinement,  case (ii) is C-confinement.   The distinction is that in S$_c$-confinement, gauge-invariant physical states containing isolated color charges are associated with an energy proportional to the separation, and this cost in energy can only be eliminated by a string breaking process which essentially neutralizes the formerly isolated color charges by binding them to other particles.  In C-confinement the energy of states with separated charges need not rise linearly, even without the intervention of a string-breaking process.
  
   In an S$_c$ confining theory, states $\Psi_V$ are inevitably metastable for large charge separation, evolving (in Euclidean time) into two
color neutral objects by string-breaking.  But the point is that a string-broken state is not a state of separated color charge; color-electric gauge fields do not emanate from color neutral objects.  The idea underlying S$_c$ confinement is to focus on the subclass of states, metastable or not, which {\it do} correspond to separated color charges, and these must be sources, because of the Gauss law, of some extended gauge field.
   
       In \cite{Greensite:2017ajx} we showed that S$_c$-confinement does not exist everywhere in the $\b-\g~$ plane of SU(2) gauge-Higgs theory, by constructing $V$ operators which do not satisfy the S$_c$-confinement criterion for sufficiently large $\g$. But this leaves open the question of whether the S$_c$ condition is satisfied {\it anywhere} in the gauge-Higgs phase diagram, apart from the pure-gauge theory
at $\g=0$.  In the next section we will show that S$_c$-confinement exists in some  $\g>0$ region of the phase diagram, and this in turn
implies the existence of a transition line between the C- and S$_c$-confinement phases, which we may speculate is identical to the
symmetry-breaking transition discussed in the previous sections.  The $V$ operators introduced in \cite{Greensite:2017ajx} have found
C-confinement only in some region above the gauge-invariant transition line shown in Fig.\ \ref{phase}.  Our conjecture is that there is
no $V$ operator which will find C-confinement below that symmetry-breaking transition line.

\section{S$_c$-confinement at strong couplings}

    We will show in this section, using strong-coupling expansions and a theorem from linear algebra, that S$_c$-confinement exists in the SU(2) gauge-Higgs system of eq.\ \rf{Sgh} if the following conditions are satisfied:
\beq
             \tg \ll \tb \ll 1 ~~~,~~~ \g \ll {1\over 10} \ ,
\label{conditions}
\eeq
where we have defined
\beq
           \tb \equiv {\b \over 4} ~~~,~~~ \tg \equiv {\g \over 4} \ .
\eeq
It should be stressed that this is an ``if'' but not an ``only if'' statement; it may be that S$_c$-confinement exists even if these
conditions are not satisfied.

\begin{figure}[h!]
\subfigure[~]  
{   
 \includegraphics[scale=0.5]{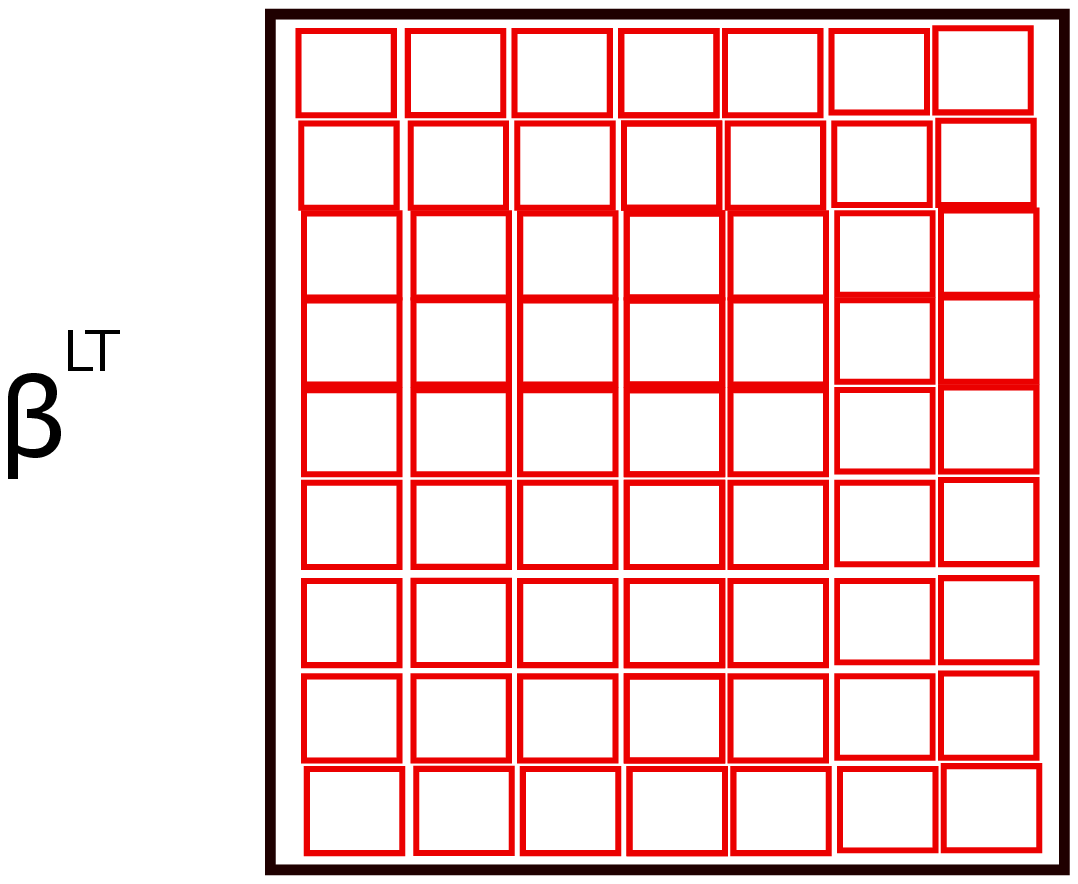}
}
\subfigure[~]  
{   
 \includegraphics[scale=0.5]{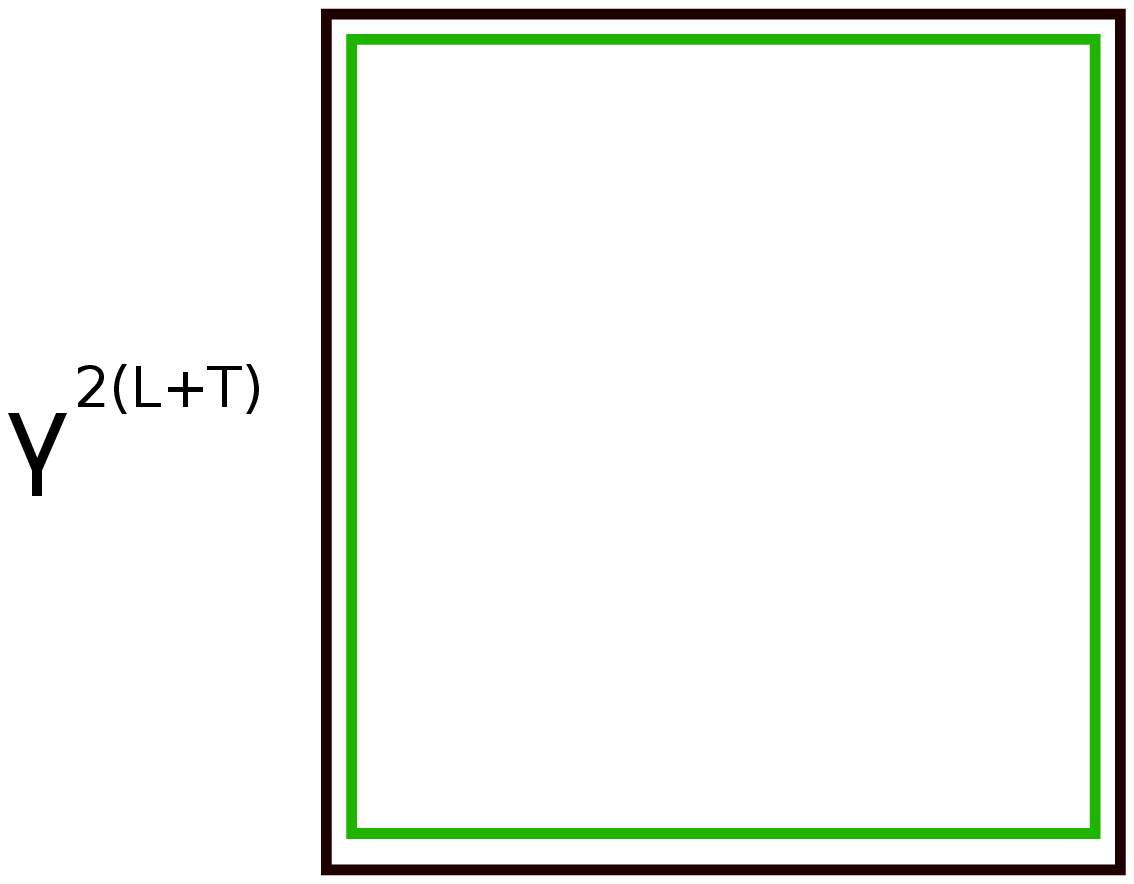}
}
\caption{Diagrammatic contributions to a rectangular $L\times T$ Wilson loop.  (a) leading order in $\b$;
(b) leading order in $\g$.} 
\label{Wloop}
\end{figure}

    In order to introduce static quark-antiquark sources at points $\vx,\vy$, we include the hopping terms
\bea
        & &   \m \sum_t \bigg\{ \overline{q}(\vx,t+1) U_0(\vx,t) q(\vx,t) \non \\
        & & \qquad + \overline{q}(\vy,t+1) U_0(\vy,t) q(\vy,t) + \mbox{h.c.} \bigg\}
\eea
in the gauge-Higgs action.  The central idea is to show that Higgs part of the action is negligible in the expression
\beq
        W_V(T) =  \langle \Psi_V | e^{-HT} | \Psi_V \rangle \ ,
\eeq
providing the conditions \rf{conditions} hold, and $T$ is small enough.  This implies that the energy expectation value, which is
the logarithmic time derivative of $W_V(T)$, will conform to the S$_c$ confinement criterion.   As a trivial example, which nonetheless illustrates the general idea, let $\vx, \vy$ be points separated by a distance $L$ along the $x$-axis, and let the operator $V(\vx,\vy,A)$ be the Wilson line running along the $x$-axis between these two points.  Then $W_V(T)$ is proportional to the expectation value $W(L,T)$ of a rectangular Wilson loop of sides of length $L$ and $T$.  The strong-coupling diagrams to leading order in $\b$ alone, and in $\g$ alone, are shown in Fig.\ \ref{Wloop}, and their contribution to $W(L,T)$ is
\beq
            2\tb^{LT} + 2\tg^{2(L+T)} \ .
\label{simpleloop}
\eeq
It is easy to see that for $L \gg T$ the $\g$ contribution is negligible compared to the $\b$ contribution providing
\beq
          T \ll  2{\log\tg  \over \log\tb} \ ,
 \eeq
 and in this limit the lattice version of the logarithmic time derivative reveals a linearly rising energy expectation value
 \beq
             E \approx - \log\left[{ W(L,T) \over W(L,T-1)}\right]
                 = (-\log \tb) L \ .
\label{logtime}
\eeq
Conversely, at times $T \gg 2\log\tg / \log\tb$, it is the $\b$ contribution that is negligible, and the energy
\beq
             E \approx -2 \log(\tg)
\eeq
is independent of separation $L$.  In other words, around time $T = 2\log\tg / \log\tb$ the string breaks, and the static charges are
screened by scalar particles.

    String-breaking is generic in gauge-Higgs theories, but the point which is illustrated in this simple example is that for $\g \ll \b$
this process takes time, which means that the energy $E_V$ of the state $\Psi_V$, which corresponds to the logarithmic time
derivative \rf{logtime} at $T=1$, obeys the S$_c$ confinement criterion.\footnote{In fact we see from \rf{simpleloop} that $E_V \propto L$
at large $L$ even at $\b=0$.  In that case we have string breaking for any $T>1$.}   This fact is unsurprising for
$V=$ a Wilson line.  The question is whether that same result is obtained for {\it arbitrary} choices of $V(\vx,\vy,A)$ which, we recall, can depend only on gauge link variables on a timeslice, and not on the matter field.  We therefore
consider the more general expression
\bea
    W_V(L,T)  &=&  \langle \Psi_0 | (\overline{q}^c(\vx) V^{cd}(\vx,\vy;A) q^d(\vy) )^\dg_{t=T}   \non \\
                      & &   e^{-HT} (\overline{q}^a(\vx) V^{ab}(\vx,\vy;A) q^b(\vy))_{t=0} | \Psi_0 \rangle \ .
\eea
After integrating out the static quark fields we have
\bea
    W_V(L,T) &=&\m^{2T} {1\over Z} \int DU D\p \tr[ V(\vx,\vy;A)_{t=0} P(\vy,T) \non \\
    & & \qquad V^\dg(\vx,\vy;A)_{t=T} P^\dg(\vx,T)] e^{-S} \ ,
\eea
where
\bea
          P(\vx,T) &=& \prod_{t=0}^{T-1} U_0(\vx,t)  
\eea 
is a timelike Wilson line.   This expression for $W_V(L,T)$ can be written as
\bea
 W_V(L,T) &=& \m^{2T}  \int DU_1 DU_2 V^{\dg cd}(U_2) \M_T^{cd,ab}(U_2,U_1) V^{ab}(U_1)  \non \\
                 &=&  \m^{2T} (V|\M_T|V) \ ,
\eea
where $V(U) = V(\vx,\vy,A)$, and
\bea
& & \M_T^{cd,ab}(U_2,U_1) =  {1\over Z} \int DU D\p \bigg\{ \prod_{\vz}  \prod_{k=1}^3 \d[U_k(\vz,0)-U_{k,1}(\vz)] \non \\  
          & & \qquad \times \d[U_k(\vz,T)-U_{k,2}(\vz)]  \bigg\}  P^{bc}(\vy,T)  P^{\dg da}(\vx,T) e^{-S}  \ .
\eea

    It is useful to introduce a basis for $V(U)$.   First define upper-lower index notation for a link variable in representation $j$
\bea
           [U^{(j)}_\m(x)]_{ab} &=&   [U^{(j)}_\m(x)]_a^{~ b} \ ,
\eea
\bea
          [U^{(j)}_\m(x)]^\dg_{ab} &=&   [U^{(j)}_\m(x)]_{~a}^b \ .
\eea
We define a cluster $\C$, in a three-dimensional time-slice, as:
\begin{enumerate}
\item a set of space-like links $\L$, connected in the sense that there is a path on the lattice between any two links in $\L$ which is
contained entirely in $\L$;
\item an SU(2) representation $j(l)=\oh,1,{3\over 2},2,...$ at each link $l \in \L$;
\item a set of vertices $\V$, and a ``color connection array'' $B$ at each vertex.  A vertex is a site shared by two or more (up to six) links 
in $\L$.  The upper and lower indices of the corresponding connection array $B_{a_1...a_n}^{b_1...b_m}(x)$ are contracted with the lower and upper indices of link variables (in various representations) which transform at that site, such that the product is a gauge singlet at the vertex $x$.
\end{enumerate}
In an SU($N>2$) gauge theory it would also be necessary to specify an orientation, i.e.\  a choice of $U_\m$ or $U^\dg_\m$ at each
link.  This choice is not strictly necessary for the SU(2) group because of the pseudoreality property of SU(2) group representations,
and inclusion of all orientations would constitute an over-complete basis.  However, any given cluster may be represented more compactly, meaning with a simpler set of connection arrays, by using a particular choice of orientations.  The simplest example of a connection
matrix at site $x$ is one which connects the color indices of a single $j=\oh$ ``ingoing'' link with one $j=\oh$  ``outgoing'' link
\beq
           [U_\m(x-\hat{\m})]_a^{~ b} B^c_b(x) [U_\n(x)]_c^{~ d}(x)   ~~~,~~~  B^c_b(x) =  {1\over \sqrt{2}} \d^c_b \ .
\eeq
If both links were represented as ingoing we would have
\beq
       [U_\m(x-\hat{\m})]_a^{~ b} B_{bc}(x) [U_\n(x)]_{~d}^c(x)   ~~~,~~~  B_{bc}(x) =  {1\over \sqrt{2}} \epsilon_{bc} \ .
\eeq
These combinations could in principle occur in equivalent representations of the same cluster.  

    In general, under a transformation $g \in SU(2)$, we have that, for states in representation $j$
\bea
            \vph(j)_a \ra  \vph'(j)_a &=& G(j)_a^{~b} \vph(j)_b \non \\
             \psi(j)^a \ra  \psi'(j)^a &=& G(j)^a_{~b} \psi(j)^b \ .
\eea
 where $G(j)$ is the gauge transformation corresponding to $g$ in representation $j$, and suppose that we have a set of $\vph,\psi$ transforming in this way. Then the connection array has the property that
\bea
            & & \vph_{1a_1}(j_1) \vph_{2a_2}(j_2)...\vph_{na_n}(j_n) \psi_1^{b_1}(j'_1) \psi_2^{b_2}...\psi_m^{b_m}(j'_m) \non \\
             & & \qquad    \times B^{a_1a_2...a_n}_{b_1 b_2...b_m}(\{j,j'\})
\eea
transforms like a singlet.  In general there may be more than one singlet in the decomposition of a product of representations, so we
distinguish among them (suppressing the dependence on the representations $\{j,j'\}$) by an additional index $\k$ in 
$B^{a_1a_2...a_n}_{b_1 b_2...b_m}(x,\k)$, with $B$ normalized such that
\beq
B^{a_1a_2...a_n}_{b_1 b_2...b_m}(x,\k)  B_{a_1a_2...a_n}^{b_1 b_2...b_m}(x,\k') = \d_{\k \k'} \ .
\label{BB}
\eeq 
 Then the gauge invariant functional $U(\C)$ defined on the cluster $\C$ is
\beq
U(\C) = \N \prod_{l=(x,k) \in L} \sqrt{2j_l+1} U^{(j_l)}_k(x) \prod_{x \in \V} B^{a_1...a_n}_{b_1...b_m}(x,\k_x) \ ,
\eeq
where $\N$ is some overall normalization constant, and it is understood that the lower(upper) indices of $B$ contract with the upper(lower) indices of links entering (leaving) a site. With these definitions, taking account of \rf{BB} and 
\beq
            \int dU  [U^{(j)}_\m(x)]_a^{~ b} [U^{(j')}_\m(x)]_{~d}^c = {\d_{jj'}\over 2j+1}\d^c_a \d_d^b
\eeq
we have
\beq
            (U(\C')|U(\C)) = \d_{\C \C'} \ .
\eeq
The simplest cluster is a Wilson loop $W_j(C)$ in representation $j$.

       A bicovariant function $U^{ab}(\C_{xy})$ is a function of links on a cluster $\C_{xy}$ which transforms like $V^{ab}(x,y;A)$.
This means that in a cluster $\C_{xy}$ there is either one single link entering or leaving sites $x,y$, or else a connection array
at $x,y$ forms a fundamental representation out of links attached to that site, rather than a singlet.   The simplest example
is a Wilson line running between sites $x$ and $y$.

\begin{figure}[htb]
\subfigure[~$M_\b(C,C)$]  
{   
 \label{Mbeta}
 \includegraphics[scale=0.35]{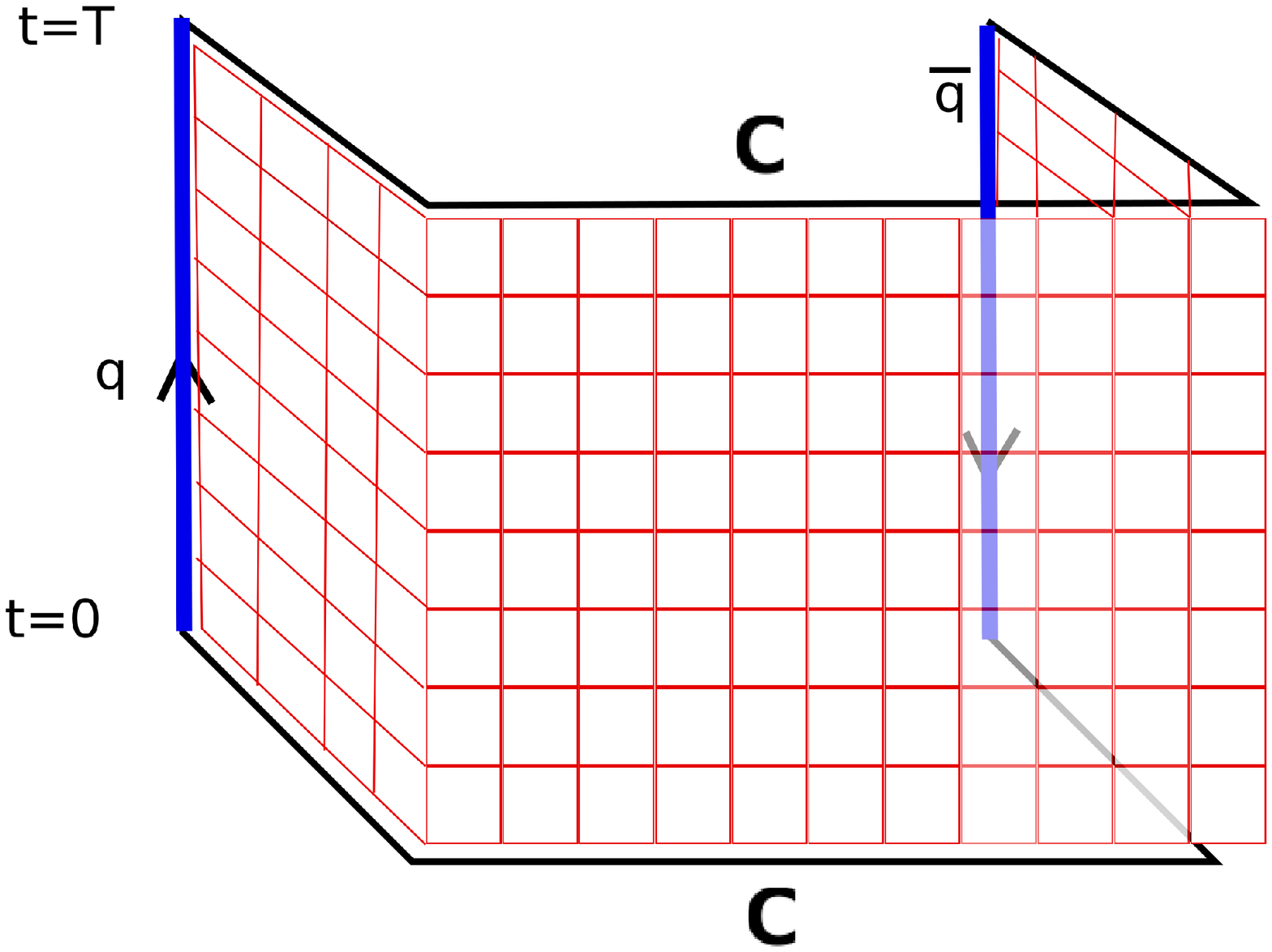}
}
\subfigure[~$M_\g(C_2,C_1)$]
{   
 \label{Mgamma}
 \includegraphics[scale=0.3]{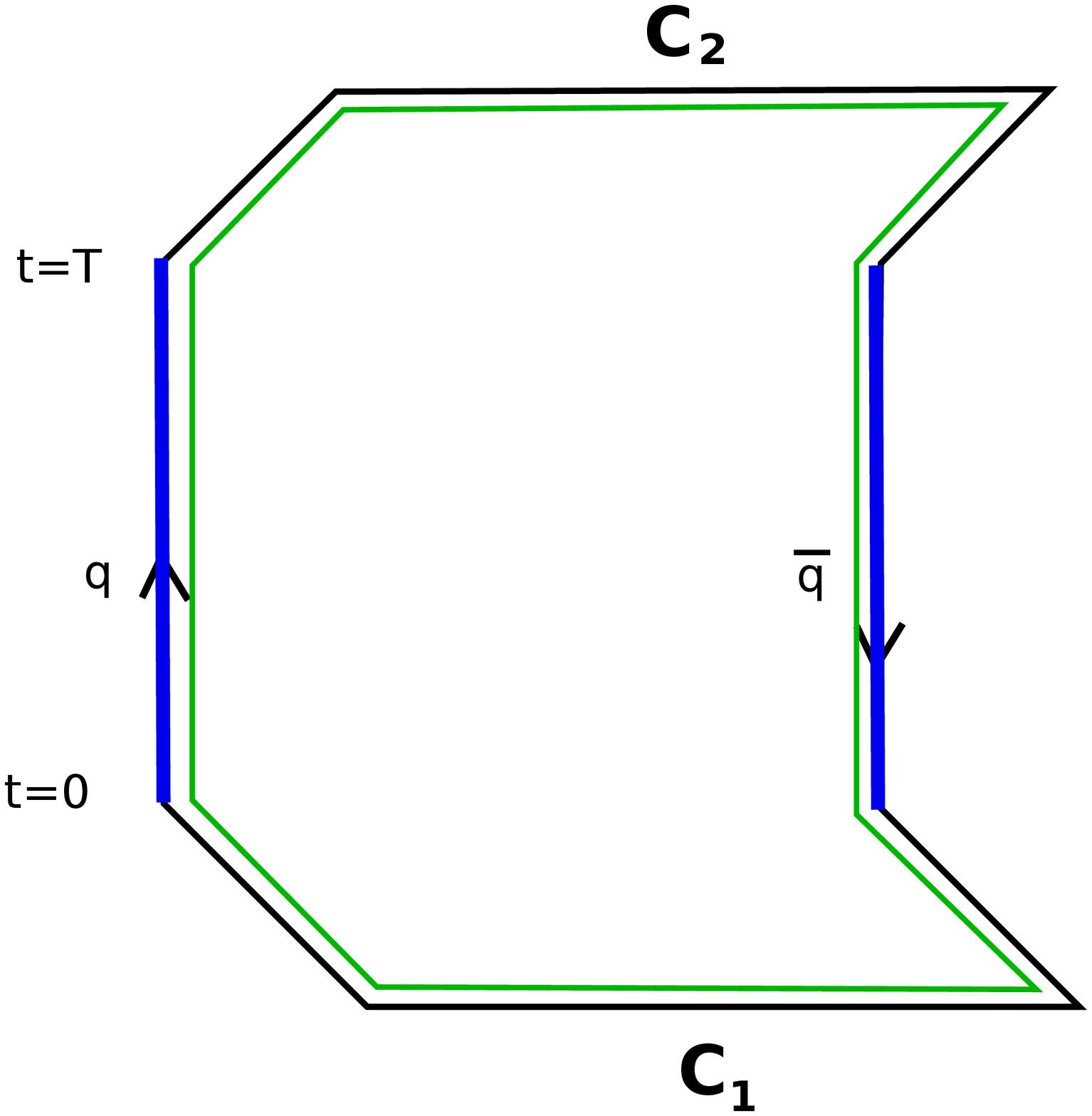}
}
\caption{(a) Diagonal contributions to $M_\b(C_2,C_1)$.  (b) Screening contributions to off-diagonal terms in $M_\g(C_2,C_1)$.}
\label{Mbg}
\end{figure}

\begin{figure}[htb]
\subfigure[~$M_{mix}(C_2,C_1)$]
{   
 \label{Mmix}
 \includegraphics[scale=0.35]{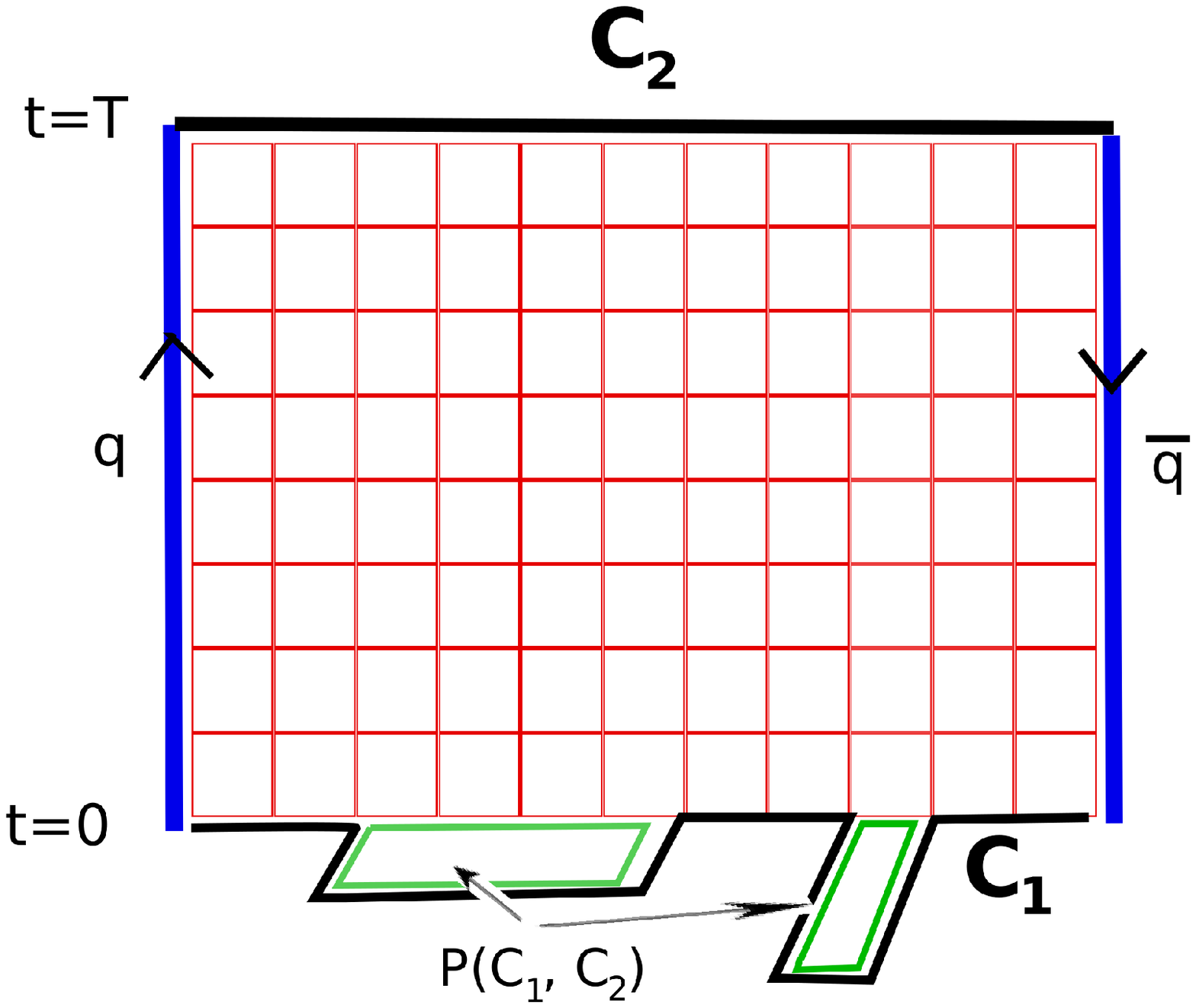}
}
\subfigure[~$B^2$ contributions]
{   
 \label{MB2}
 \includegraphics[scale=0.35]{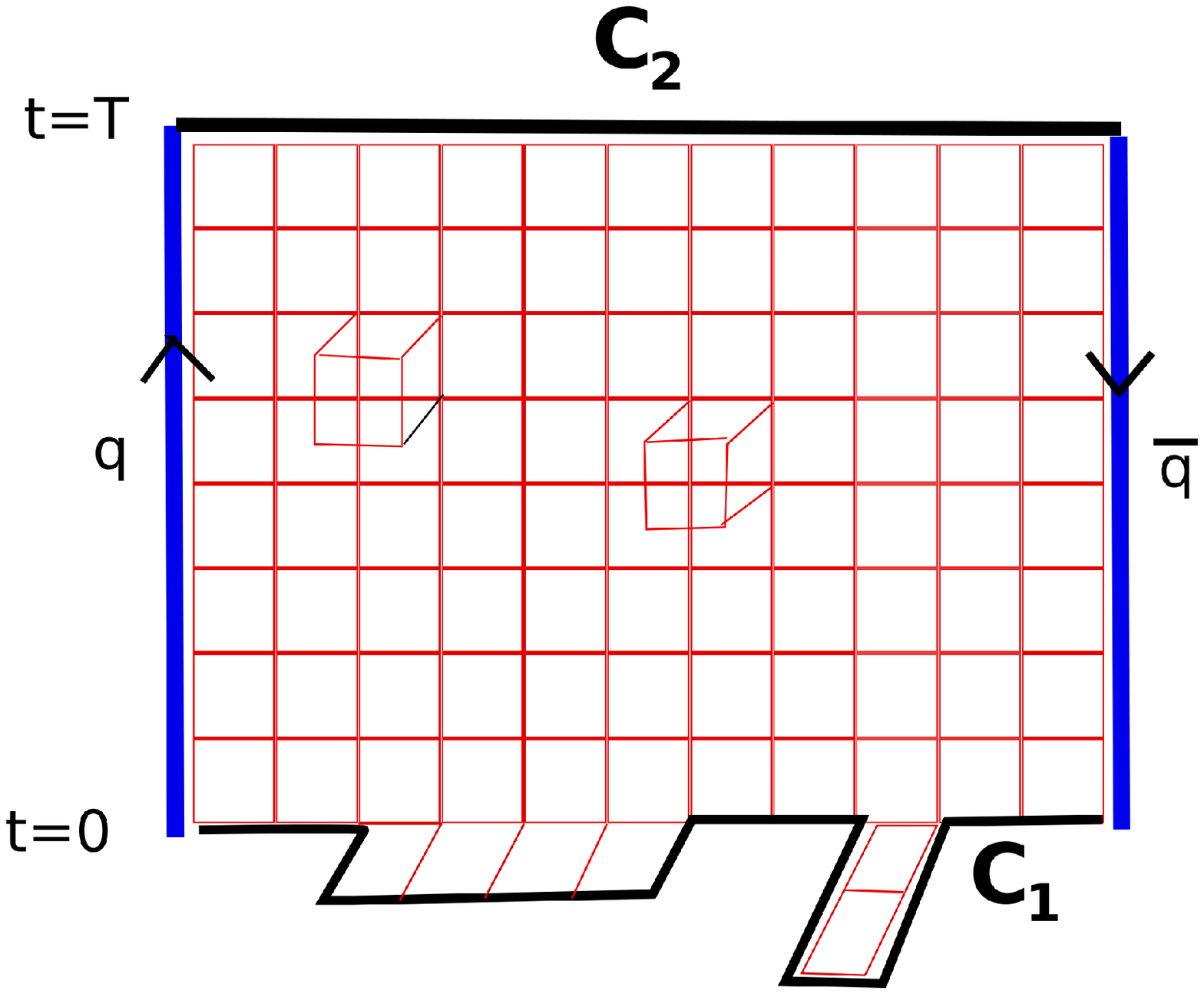}
}
\caption{(a) Off-diagonal ``confinement-type'' contributions to $M_{mix}$.  (b) off-diagonal contributions to $M_\b$.  These necessarily
include $B^2$ terms.} 
\label{mixture}
\end{figure}

         We now define the matrix in a cluster basis
\bea
  \lefteqn{M_T(\C_{2xy},\C_{1xy})  =}   \\ 
  & & \int DU_2 DU_1   U^{\dg cd}(\C_{2xy}) \M_T^{cd,ab}(U_2,U_1) U^{ab}(\C_{1xy}) \ ,
\eea
and the strategy is to estimate the largest eigenvalues of this matrix.  The logarithmic lattice time derivative of the largest eigenvalue
$\l_{max}(T)$ is simply $-\log [\l_{max}(T)/\l_{max}(T-1)]$, and the case of $T=1$ is of particular interest, since in this case
$\l_{max}(1)$ is the largest eigenvalue of the transfer matrix restricted to the subspace of $\Psi_V$ states.  Then
\beq
          E_V \ge E_{min} = -\log \l_{max}(1) \ ,
\eeq
and we have S$_c$ confinement if  $E_{min}$ is bounded from below by a function which rises linearly with  $L_0=|x-y|$.   In an S$_c$ confining theory, a $\Psi_V$ state is metastable, and evolves in Euclidean time to a lower energy state consisting of two color neutral objects via
the usual string breaking process.  We stress again that stability is in no way a condition for S$_c$ confinement.  String breaking takes place when $T$ is large enough such that $-\log [\l_{max}(T)/\l_{max}(T-1)]$ ceases to rise linearly with separation $L_0$.

\subsubsection{Simple contours and the Gershgorin Theorem}

We begin with the simplest possible bicovariant clusters, i.e.\
Wilson lines in the fundamental ($j=\oh$) representation, running along contours $C_1, C_2$ between points $x$ and $y$.  Within the
strong coupling expansion, with $\tg \ll \tb \ll 1$ and small $T$, it can be seen that the {\it leading} contributions
to each matrix element $M(C_2,C_1)$ are either ``confining'', in the sense that their logarithmic time derivatives are greater than or equal
to $-\log(\b) L'$, where $L' \ge |\vx-\vy|$, or else they are ``screening'', in the sense that the logarithmic time derivative is a constant independent of  $L_0=|\vx-\vy|$.  The diagonal elements $M(C_2,C_1)$ with
$C_2=C_1$, and also ``near-diagonal'' elements, are of the confining type, while most off-diagonal elements are of the screening type.
Since the off-diagonal screening matrix elements are far more numerous than confining matrix elements, the question is whether these
screening elements can overwhelm the confining elements, and result in a ground state with non-confining behavior.

\begin{figure*}[t!]
\subfigure[~$M_{\b}(C_2,C_2)$]  
{   
 \label{Mbb}
 \includegraphics[scale=0.2]{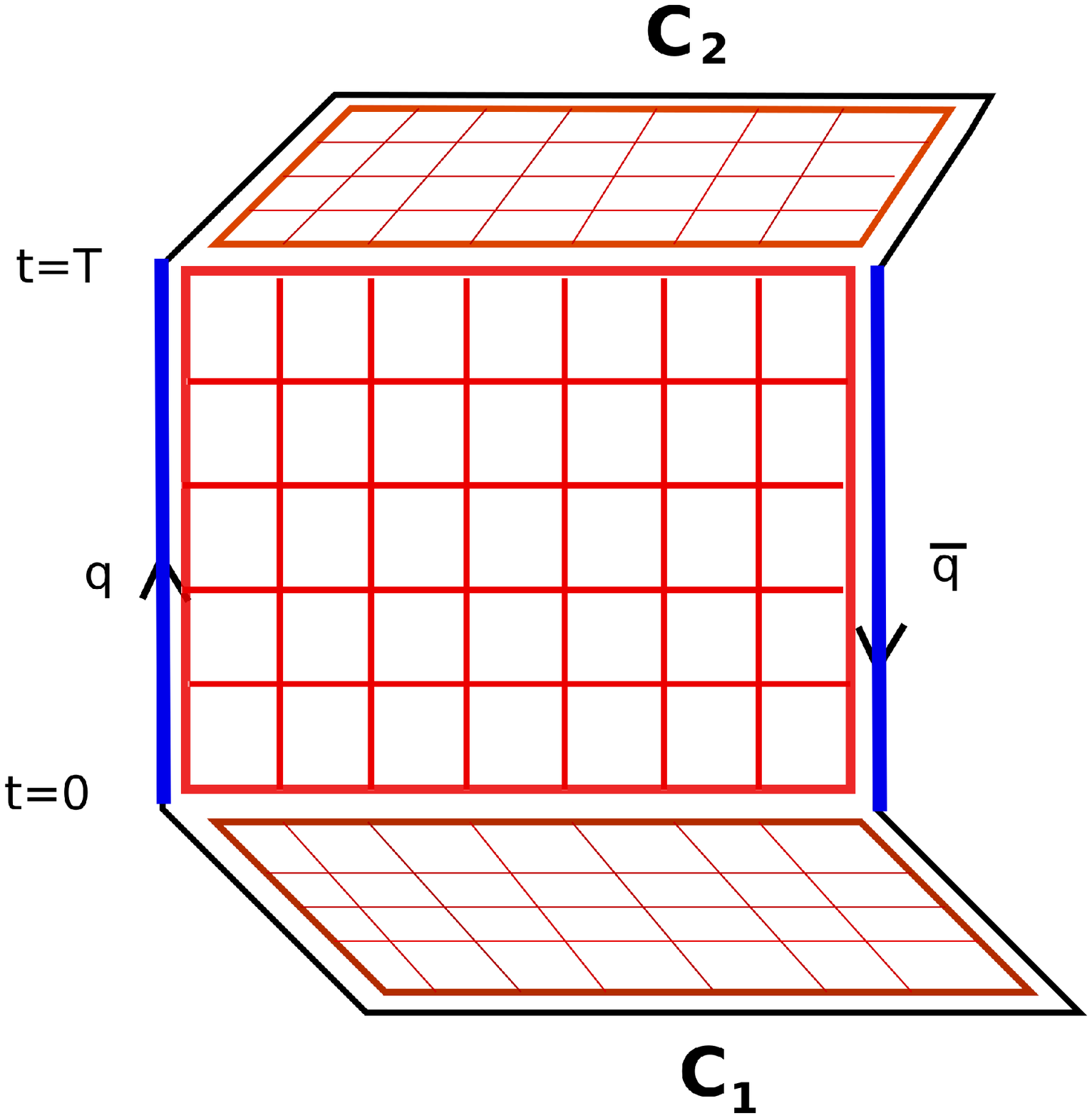}
}
\subfigure[~$M_{mix}(C_2,C_1)$]  
{   
 \label{Mmix2}
 \includegraphics[scale=0.2]{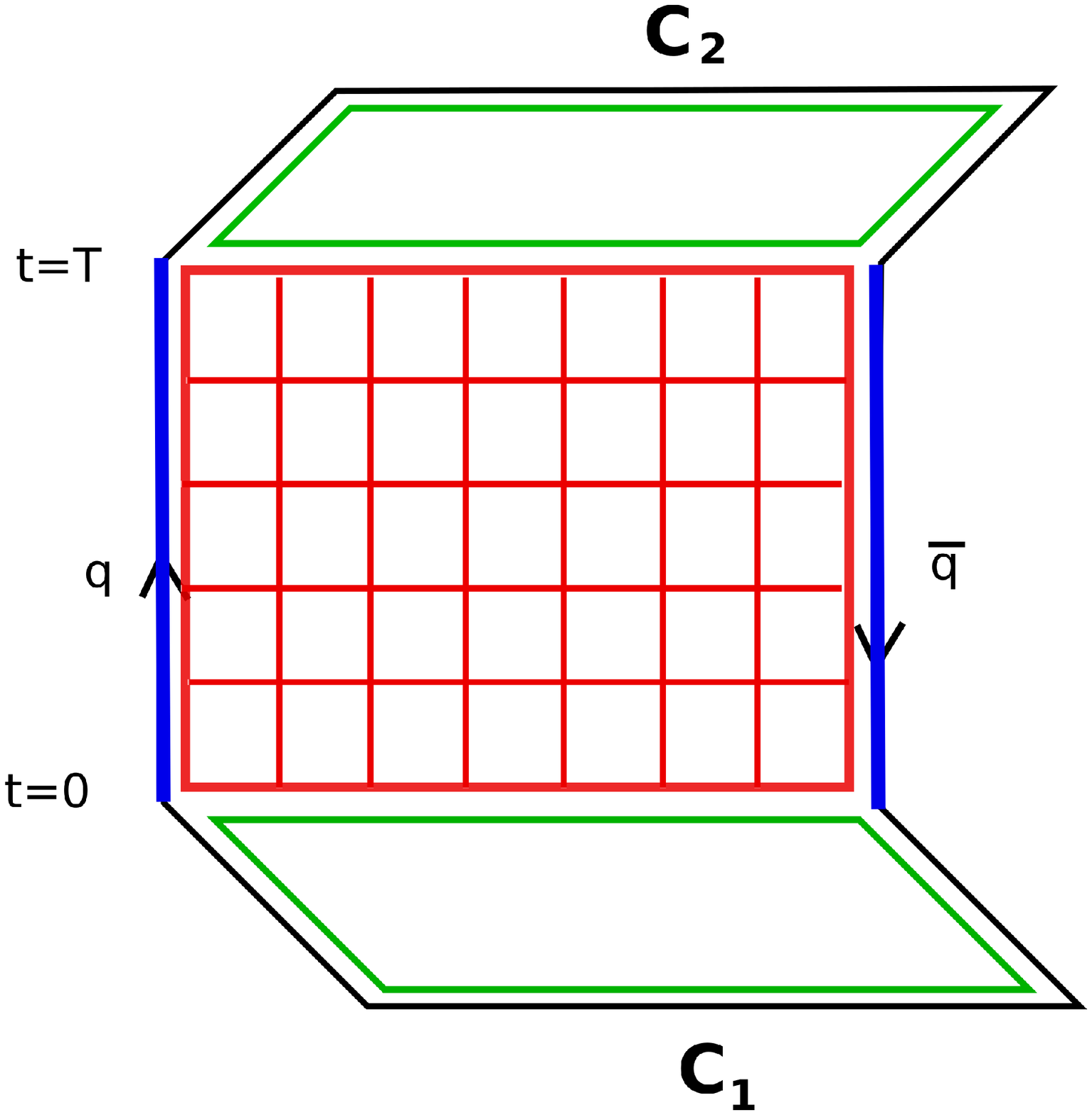}
 }
\caption{A comparison of off-diagonal confinement-type contributions in (a) $M_\b$ and (b) $M_{mix}$. }
\label{bbmix}
\end{figure*}

      The gauge-Higgs action consists of an ``$E^2$-term,'' which is the sum over timelike plaquettes in the Wilson action, a ``$B^2$-term,"\footnote{Not to be confused, of course, with the $B$ connection arrays.}
which is the sum over spacelike plaquettes in the Wilson action, and a Higgs-term, which is the term proportional to $\g$.  It simplifies the
discussion to initially ignore the $B^2$ term.  Then the leading contributions, under the condition that $T \ll \log\g/\log\b$, can be grouped into terms that depend only on $\b$, which are confining, that depend only on $\g$, which are screening, and mixed terms that depend on both 
$\b$ and $\g$, i.e.:
\bea
         M_T(C_2,C_1)  &=&  M_\b(C_1,C_1)\d_{C_2,C_1} + M_\g(C_2,C_1) \non \\
                              & & + M_{mix}(C_2,C_1) \ ,
\eea
where the leading contributions in $M_\b, M_\g$ go as
\bea
         M_\b(C,C) &=& \tb^{L(C)T}  \label{beta} \\
         M_\g(C_2,C_1) &=& \tg^{L(C_1)+L(C_2)+2T} \label{gamma} \ ,
\eea
If loops $C_1,C_2$ are large in the sense that 
\beq 
\b^{L(C_1) T} \mbox{~~and~~}  \b^{L(C_2)T} \ll \b^{L_0 T} \g^{2L_0}  
\eeq 
then the leading contribution
to $M_{mix}$ is
\beq
           M_{mix}(C_2,C_1) =  \tb^{L_0 T} \tg^{L(C_1)+L(C_2)+2L_0+ 2T} \ ,
\label{mix0}
\eeq
otherwise
\bea
 M_{mix}(C_2,C_1) &=& \tb^{L' T} \tg^{P(C_1,C_2)}  \ ,
 \label{mix}
\eea 
where $L_0$ in \rf{mix0} is the minimal distance between points $\vx$ and $\vy$, $L'$ in \rf{mix} is the smaller of $L(C_1)$ and $L(C_2)$, and $P(C_1,C_2)$ is the perimeter of the area enclosed by curves $C_1,C_2$ ( which is $ \le L(C_1)+L(C_2)$). Examples of the diagrammatic representation of these matrix elements are shown in Fig.\ \ref{Mbeta} for \rf{beta}, Fig.\ \ref{Mgamma} for \rf{gamma}, Fig.\ \ref{Mmix2} for \rf{mix0} and Fig.\ \ref{Mmix} for \rf{mix}.  The $M_{mix}$ confining matrix element is negligible compared to the corresponding $M_\g$ matrix element for $L_0 \gg T$, $P(C_1,C_2) \approx L(C_1)+L(C_2)$, and
when
\beq
          T \ll  {\log\tg  \over \log\tb} \ .
\label{T}
\eeq  

   Now consider the matrix
\beq
           \widetilde{M}(C_2,C_1)  =  M_\b(C_1,C_1)\d_{C_2,C_1} + M_\g(C_2,C_1) \ .
\eeq
This is a matrix of enormous dimensionality, and only the diagonal terms are confining.  The rest, i.e.\ the vast majority, are screening.
So the question is whether the eigenvalues of this matrix are of the confining or screening type.  This question can be answered with
the help of the following theorem (see e.g. \cite{Golub}):

\bigskip

\noindent\fbox{
    \parbox{8cm} {

\ni \underline{\bf The Gershgorin Circle Theorem} \\

Let $A$ be a complex  $n\times n$ matrix, with matrix elements $A_{ij}$, and let
\beq
         r_i = \sum_{j \ne i} |A_{ij}|
\eeq
be the sum of the magnitudes of the off-diagonal entries in the  $i$-th row.  Let $D_i$ be a closed disk of radius $r_i$, centered at
$A_{ii}$, in the complex plane.  These, for $i=1,2,...n$, are known as  ``Gershgorin disks."  The theorem states that every eigenvalue of $A$ must lie within at least one of the Gershgorin disks. In particular, let an eigenvalue $\l$ correspond to an eigenvector $\vec{u}$ with
$u_i = 1$ and $|u_j| < 1$ for all $j \ne i$.  Then
\beq
           |\l - A_{ii}| \le r_i \ .
\eeq 
           }     }
          
\bigskip

    With this motivation we compute an upper bound for $r_C$:
\bea
             r_{C} &=& \sum_{C_1 \ne C} \widetilde{M}(C,C_1)  \non \\
                      &\approx& \tg^{L(C)+2T} \sum_{C_1 \ne C}  \tg^{L(C_1)}  \non \\
                      &<& \tg^{L(C)+2T}  \sum_{L=L_0}^\infty \tg^L N(L) \ ,
\eea
where $N(L)$ is the number of open contours with endpoints $\vx,\vy$ of length $L$.  Without the second endpoint restriction,
$N(L) = 5^L$ since at each step there are five possible directions to go without backtracking.  This will serve as an upper limit
\bea
          r_{C} &<&  \tg^{L(C)+2T}  \sum_{L=L_0}^\infty (5 \tg)^L \non \\
                    &<& \tg^{L(C)+2T}  {(5 \tg)^{L_0} \over 1 - 5 \tg} \ ,
\eea
and this radius should be compared with the diagonal term $M_\b(C,C)$ in \rf{beta}.  Assuming the conditions \rf{conditions} are satisfied, and that $T$ is small enough so that \rf{T} also holds, then the radius $r_C$ of the Gershgorin disk $D_C$ is negligible compared to the diagonal term $\widetilde{M}(C,C)$.   Since also $M_{mix}(C_2,C_1)\ll M_\g(C_2,C_1)$, it follows that the 
difference between the largest eigenvalues of $M_T$ and the corresponding eigenvalues of $M_\b$ are, by the Gershgorin theorem, 
negligible.  This in turn implies S$_c$ confinement.

   Adding back the $B^2$ term in the action does not change this conclusion.  The effect of the $B^2$ term is to introduce subleading dressings of the sheet of plaquettes, and to allow for new contributions to off-diagonal elements of the confining type.  These types of contributions are both illustrated in Fig.\ \ref{MB2}.   There are certainly contours $C_1,C_2$ such that $M_{mix}(C_2,C_1)$ is larger than $M_\b(C_2,C_1)$, as
would be the case for the contributions shown in Fig.\ \ref{bbmix} if the combined area $A(C_1,C_2)$ bounded by $C_1$ and $C_2$ in a plane is such that 
\beq
             \tb^{A(C_1,C_2)} <  \tg^{L(C_1)+L(C_2)} \ .
\eeq
But inspection of such terms
(compare, e.g., Fig.\ \ref{Mmix2} with Fig.\ \ref{Mgamma}) shows that, under the assumption \rf{T}, they are always very much smaller than the corresponding matrix elements of $M_\g(C_2,C_1)$ which, we have already argued from the Gershgorin theorem, can be neglected.  So apart from negligible terms, and assuming the conditions
\rf{conditions} and \rf{T}, all of the leading terms in $M_T(C_2,C_1)$ are in $M_\b(C_1,C_2)$; i.e.\ the terms of a pure gauge theory.  Then the eigenvalue spectrum 
of $M_T$, for times $T$ up to the limit in \rf{T}, will not differ much from that of a pure gauge theory.
   
     It follows that up to this limit in $T$, the spectrum of static $q\overline{q}$ states, obtained by evolving states of the form $\Psi_V$ shown in \rf{Vstate} for Euclidean time $T$, will have energies bounded by a linear potential, with a string tension which is closely approximated by that of a pure gauge theory.  Beyond that time, screening will take over.   This
is S$_c$-confinement, but the argument assumes that we restrict the basis states to simple open contours. 

\subsubsection{Summation over clusters}

    So the next step is to enlarge the basis and, motivated by the Gershgorin theorem, we consider the sum over general clusters in the screening terms
\beq
 r_{C} = \sum_{\C_1 \ne C} |M_\g(C,\C_1)| \ .
\eeq
It simplifies matters at this point to go to unitary gauge ${\phi(x)=\mathbbm{1}}$.   
Since all group representations are in play, we will need the SU(2) character expansion
\beq
  \exp\left[\oh \g \tr[U]\right] = \sum_j c_j(\g) \chi_j[U] \ ,
\label{form1}
\eeq
where
\bea
c_j(\g) &=& 2(2j+1) {I_{2j+1}(\g) \over \g} \non \\
           &\approx& {2j+1 \over (2j+1)!} \left({\g \over 2}\right)^{2j} \ .
\label{cj}
\eea
We then have for the off-diagonal terms in $M_\g$, to lowest order in $\g$,
\begin{widetext}
\bea
M_\g(C,\C_1) &=&  \tg^{L(C)+2T}  \int DU  \prod_{l\in \L} \sqrt{2j_l+1} U^{(j_l)}(l) \prod_{x \in \V} B^{a_1...a_n}_{b_1...b_m}(x,\k_x) 
    \prod_{l'\in \L} \sum_{j_l'} c_{j_{l'}} \tr[U^{(j_{l'})}(l')] \ .
\eea   
Carrying out the $U$ integrations and using \rf{cj},
\bea
M_\g(C,\C_1) &=&  \tg^{L(C)+2T}  \prod_{l\in \L} \sum_{j_l} {\sqrt{2j_l+1} \over (2j_l+1)!} \left( \g\over 2\right)^{2j_l}  \left(\prod_{x \in \V} B^{a_1...a_n}_{b_1...b_m}(x,\k_x)\right)_{contracted}
\eea 
\end{widetext}
Each upper (lower) index in a $B$ array at vertex site $x$ is associated with a lower (upper) index on a link attached to that site.
The meaning of $(\prod B)_{contracted}$ is that each upper (lower) index in $B(x)$ associated with a particular link is contracted with the corresponding lower (upper) index associated with the same link in a $B$ array at a neighboring vertex, i.e. the pattern is
\beq
  B^{\cdot \cdot a \cdot \cdot}_{\cdot \cdot \cdot}(x) B^{\cdot \cdot \cdot}_{\cdot \cdot a \cdot \cdot}(x+\hat{k}) ~~ \mbox{or} ~~
   B^{\cdot \cdot \cdot}_{\cdot \cdot a \cdot \cdot}(x)B^{\cdot \cdot a \cdot \cdot}_{\cdot \cdot \cdot}(x+\hat{k}) \ .
\eeq
Unfortunately we have no general 
formula for $(\prod B)_{contracted}$, but we can argue for a rough upper bound based on the normalization condition \rf{BB}.  Suppose all array elements have about the same magnitude.  Then \rf{BB} requires that
\beq
|B^{a_1a_2...a_n}_{b_1 b_2...b_m}(x,\{j,j'\},\k) | \sim \prod_{i=1}^n  {1\over \sqrt{2j_i+1}} \prod_{k=1}^m  {1\over \sqrt{2j'_k+1}} \ ,
\label{constB}
\eeq
where the products are over links entering and leaving vertex $x$.  In that case it is easy to see that $(\prod B)_{contracted} \sim 1$.  In fact the assumption of equal magnitude array elements results in a large overestimate, as can be seen for the case that $\C_1$ is a simple
open contour composed of links in representation $j$, in which case
\beq
          B^{s_1}_{s_2} B^{s_2}_{s_3}...B^{s_{L}}_{s_{L+1}} =  {1 \over (2j+1)^{L/2}} \d^{s_1}_{s_{L+1}} \ .
\label{twoindexprod}
\eeq
The reason for the overestimate is that because $B^a_b = {1\over \sqrt{2j+1}} \d^a_b$, the magnitude of the ``average'' array element
is really ${(2j+1)^{-3/2}}$, rather than $(2j+1)^{-1}$.   

    In a little more generality, consider the product of a set of $B$ arrays with a total of $N$ upper indices and an equal number of
lower indices, and, initially, no sum over indices.  Each index is associated with an SU(2) representation $j$, with the index running
from 1 to $2j+1$.  Consider choosing each index at random, within its allowed range, and let 
\bea
\overline{B} &=& \sim \prod_{i=1}^n  {1\over 2j_i+1} \prod_{k=1}^m  {1\over 2j'_k+1}\sum_{a_1=1}^{2j_1+1}\sum_{a_2=1}^{2j_2+1}... \sum_{a_1=1}^{2j_n+1}\non \\
 & & \qquad  \sum_{b_1=1}^{2j'_1+1}\sum_{b_2=1}^{2j'_2+1}...\sum_{b_m=1}^{2j'_m+1} |B^{a_1a_2...a_n}_{b_1 b_2...b_m}|
\eea
be the average of the moduli
of the array elements in a given $B$ array.  Then the expectation value of the modulus of the product of this random choice of (real-valued) array elements is simply the product of average values, i.e.\
\bea
    \lefteqn{\langle | B^{\cdot \cdot \cdot}_{\cdot \cdot \cdot}(x_1) B^{\cdot \cdot \cdot}_{\cdot \cdot \cdot}(x_2)...B^{\cdot \cdot \cdot}_{\cdot \cdot \cdot}(x_n)| \rangle} \non \\
     & & \qquad = \langle | B^{\cdot \cdot \cdot}_{\cdot \cdot \cdot}(x_1)|   | B^{\cdot \cdot \cdot}_{\cdot \cdot \cdot}(x_2)|  
          ...  |B^{\cdot \cdot \cdot}_{\cdot \cdot \cdot}(x_n)| \rangle \non \\
         & &  \qquad =\overline{B}(x_1) ~\overline{B}(x_2)...\overline{B}(x_n) \ .
\label{prodB}
\eea
Now suppose, in the first line of \rf{prodB}, that we pair each upper index with a lower index such that the paired indices 
belong to different $B$'s, and assign the
same value to each paired index.  This reduces the number of indices which can be chosen randomly from $2N$ to $N$, but under
a random choice of the remaining $N$ index values the expectation value of any $|B^{\cdot \cdot \cdot}_{\cdot \cdot \cdot}(x)|$ appearing in the product is again  
$\overline{B}(x)$.  The pairing restriction introduces a weak correlation among the different 
$B$'s in the product, but if we ignore this correlation then the expectation value of the modulus of the product with paired indices is still \rf{prodB}.  If we denote the values of the $i$-th set of paired indices as ${a_i,a_i}$, and then sum over those values, we then have the estimate
\bea
       \lefteqn{\sum_{a_1=1}^{2j_1+1} \sum_{a_2=1}^{2j_2+1} ... \sum_{a_N=1}^{2j_N+1} B^{\cdot \cdot \cdot}_{\cdot \cdot \cdot}(x_1) B^{\cdot \cdot \cdot}_{\cdot \cdot \cdot}(x_2)...B^{\cdot \cdot \cdot}_{\cdot \cdot \cdot}(x_n) }
\non \\
       && \qquad \approx \left(\prod_{i=1}^{N} (2j_i+1) \right) \overline{B}(x_1) ~\overline{B}(x_2)...\overline{B}(x_n)    \ .
\eea
To support the validity of this approximation we return to the simplest case, \rf{twoindexprod}.  Since the average of the each array is
$(2j+1)^{-3/2}$, and summation over each contracted index gives a factor of $(2j+1)$, the approximation delivers the correct overall factor
$(2j+1)^{-L/2}$.

    Now, under the constraint \rf{BB}, it is easy to show that the average value $\overline{B}$ of any array
\beq
           B^{a_1 a_2 ... a_n}_{b_1 b_2 ... b_m}   ~~,~~ a_i = 1,2,...,2j_i+1 ~~,~~ b_i = 1,2,...,2j'_i+1 
\eeq
is maximized when all array elements are identical, and equal to the right hand side of \rf{constB}.
This then leads us to the upper bound 
\beq
\left|(\prod B)_{contracted} \right| \le 1  \ ,       
\eeq
and from that we obtain 
\beq
|M_\g(C,\C_1)| <  \tg^{L(C)+2T}  \prod_{l\in \L} {\sqrt{j_l+1} \over (2j_l+1)!}  \left( \g\over 2\right)^{2j_l} \ .
\eeq

    Now summing over all possible clusters, we have from the Gershgorin theorem
\bea
r_C &<& \tg^{L(C)+2T} \sum_{\L}   \prod_{l\in \L} \sum_{j_l}  {\sqrt{j_l+1} \over (2j_l+1)!}  \left( \g\over 2\right)^{2j_l} \prod_{x\in \V} n_s(x,\{j,j'\})  \ ,  \non \\ 
\eea
where $n_s(x,\{j,j'\})$ is the number of singlets (which may be zero, if the set does not form a cluster) at vertex site $x$, and this number depends on the representations 
$\{j,j'\}$ of links entering/leaving the site.  An upper bound on the number of singlets that can be formed is the number of orthogonal states that can be formed at $x$, i.e.
\beq
        n_s < \prod_i (2j_i+1) \ ,
\eeq
where the product is over each link attached to the site.  Then absorbing two factors of $2j_l+1$ (from each end of link $l$) into the product over links, 
\bea
r_C &<& \tg^{L(C)+2T}  \sum_{\L}   \prod_{l\in \L}  \sum_{j_l} { (2j_l+1)^{5/2} \over (2j_l+1)!}  \left({\g \over 2}\right)^{2j_l} \ .
\eea
and we note that
\bea
\lefteqn{\sum_{j_l=\oh,1,{3\over 2},2,...} { (2j_l+1)^{5/2} \over (2j_l+1)!}  \left({\g \over 2}\right)^{2j_l} } \non \\
                & & \qquad < \sum_{n=1}^\infty {(n+1)^3 \over (n+1)!} \left({\g \over 2}\right)^n \non \\
                & & \qquad = \frac{1}{4} \left(e^{\gamma /2} \gamma ^2+6 e^{\gamma/2} \gamma +4 e^{\gamma /2}-4\right) \non \\
                & & \qquad  = 2\g + O(\g^2) \ .
\eea
Consequently, proceeding along the lines of the previous subsection,
\bea
          r_C &<& \tg^{L(C)+2T}  \sum_{L=L_0}^\infty 5^L (2 \g)^L    \non \\
                &<& \tg^{L(C)+2T} { (10 \g)^{L_0} \over 1 - 10 \g}  \ ,
\label{bnd2}
\eea
where we recall that $L_0$ is the minimal distance on the lattice between sites $x,y$.  So this is the bound on the sum of off-diagonal terms.  The diagonal term from $M_\b$, to leading order in $\b$, is
\beq
           M_\b(C,C) = \left({\b \over 4}\right)^{L(C)T} \ ,
\eeq
which means that $r_C \ll M_\b(C,C)$ providing that $\tg \ll \tb \ll 1$ with $T$  small enough so that condition \rf{T} is satisfied, while convergence of the sum in \rf{bnd2} also requires
\beq
  \g \ll {1\over 10} \ .
\eeq

\begin{figure}[htb]
 \subfigure[~]  
{   
\includegraphics[scale=0.25]{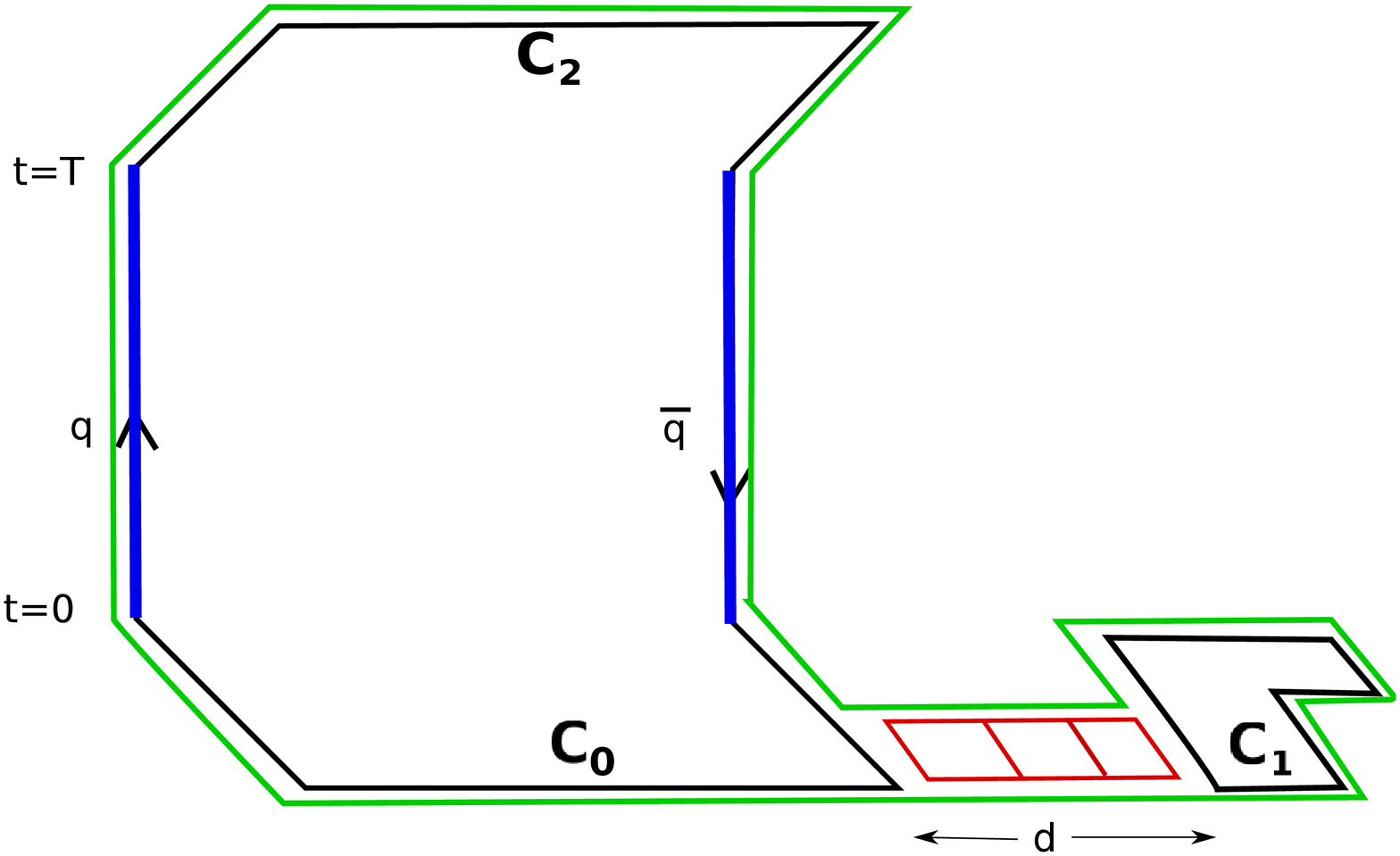}
\label{dgamma}
}
 \subfigure[~]  
{   
\includegraphics[scale=0.3]{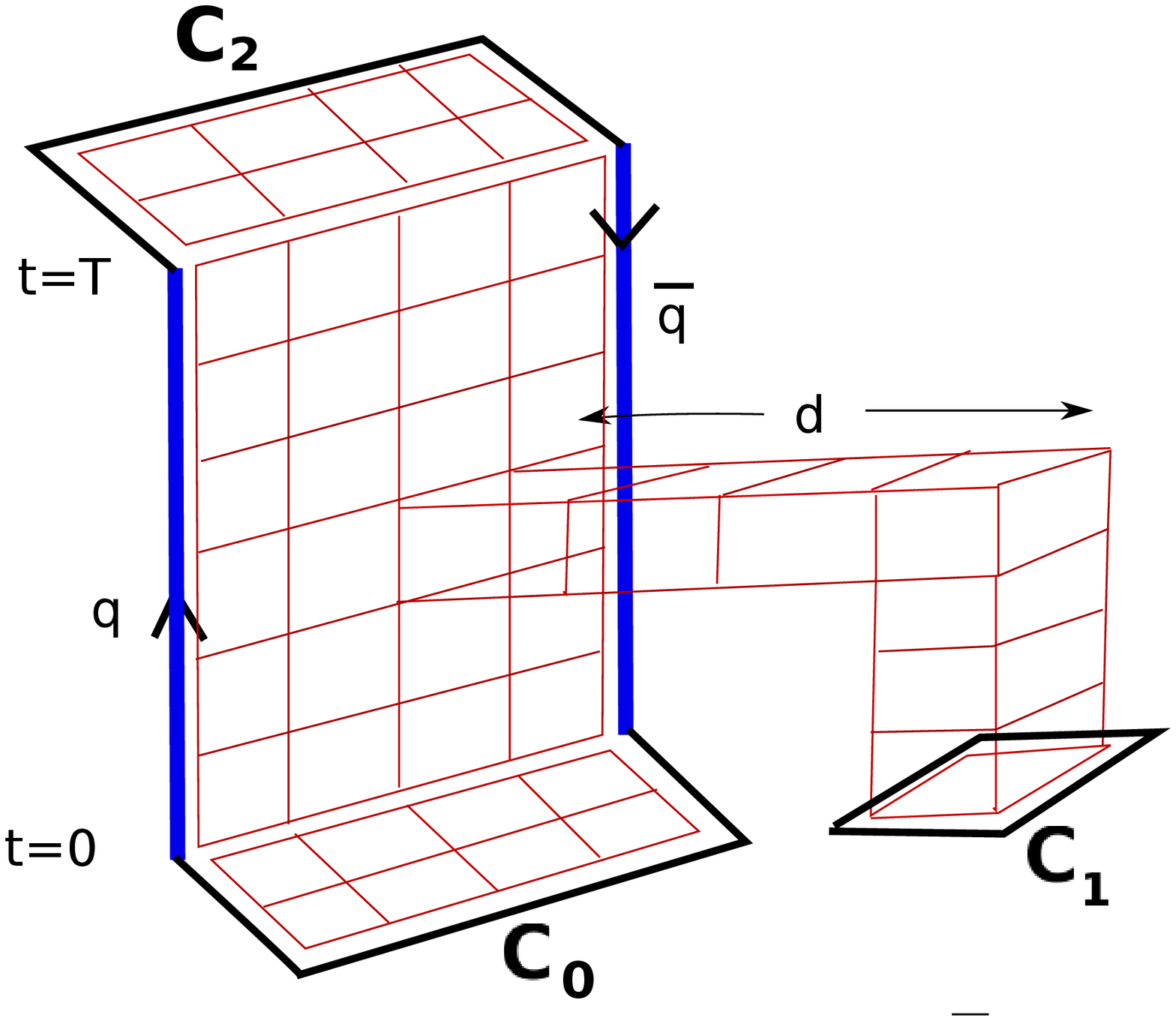}
\label{disconnect}
}
\caption{Connected diagrams for disconnected clusters in the initial state. (a) screening contributions; (b) confining contributions.}
\label{dconn}
\end{figure}  

    Of course one can go on to consider the more general case
\beq
           r_{\C_2} = \sum_{\C_1 \ne \C_2} |M_T(\C_2,\C_1| \ ,
\eeq
and, by the previous analysis, establish an upper bound
\beq
        r_{\C_2} < \tg^{2T}  \prod_{l\in \L_2} {\sqrt{j_l+1} \over (2j_l+1)!}  \left( \g\over 2\right)^{2j_l} 
         { (10 \g)^{L_0} \over 1 - 10 \g}  \ ,
\eeq
with a diagonal term
\beq 
          M_\b(\C_2,\C_2) < \left({\b\over 4}\right)^{L(\C_2)T} \ .
\eeq
But we are interested in the largest eigenvalue of $M_T$.  Let $C_{0xy}$ be the contour of minimal distance $L_0$ between $x$ and
$y$.  Then the largest diagonal term $M_T(\C,\C)$ corresponds to $\C=C_{0xy}$, while in general
$r_\C \ll M_T(C_{0xy},C_{0xy})$ for any $\C$.  It follows that the largest eigenvalue is within $r_{C_{0xy}}$ of $M_T(C_{0xy},C_{0xy})$,
which for small $T$ is a small correction to the pure gauge value.  This implies that the logarithmic time derivative of the largest
eigenvalue satisfies the S$_c$ confinement condition.

\subsubsection{Summation over disconnected clusters}

    Once we include the $B^2$ term we must also consider matrix elements such as $M_T(C,\{\C_1,\C_2,...\}]$, where $\C_{1xy},\C_2,...$ are
disjoint clusters having no links in common.  Only connected diagrams contributing to $M_T$ are relevant.   
Some simple examples of relevant strong coupling diagrams connecting such disjoint
clusters are shown in Fig.\ \ref{dgamma}, which is a contribution to $M_{mix}$, and Fig.\ \ref{disconnect},  which is a contribution to 
$M_\b$.  We are mainly concerned with contributions of the type $M_{mix}$, since off-diagonal contributions in $M_\b$ are known to be compatible with S$_c$ confinement in pure gauge theories.  The goal is to place an upper bound on the contribution to the Gershgorin disk radius $r_C$ due to the sum
\beq
           \sum_{\{\C\}_{dis} \ne C} |M_{mix}(C,\{\C\}_{dis};\b,\g)| \ ,
\label{rdis}
\eeq
where the sum is over all sets of clusters $\{\C_{1xy},\C_2,...\}$ which are disconnected in the sense that they have no links in common, and
the dependence of $M_{mix}$ on both couplings is indicated here explicitly.

\begin{figure*}[t!]
 \subfigure[~]  
 {
 \includegraphics[scale=0.2]{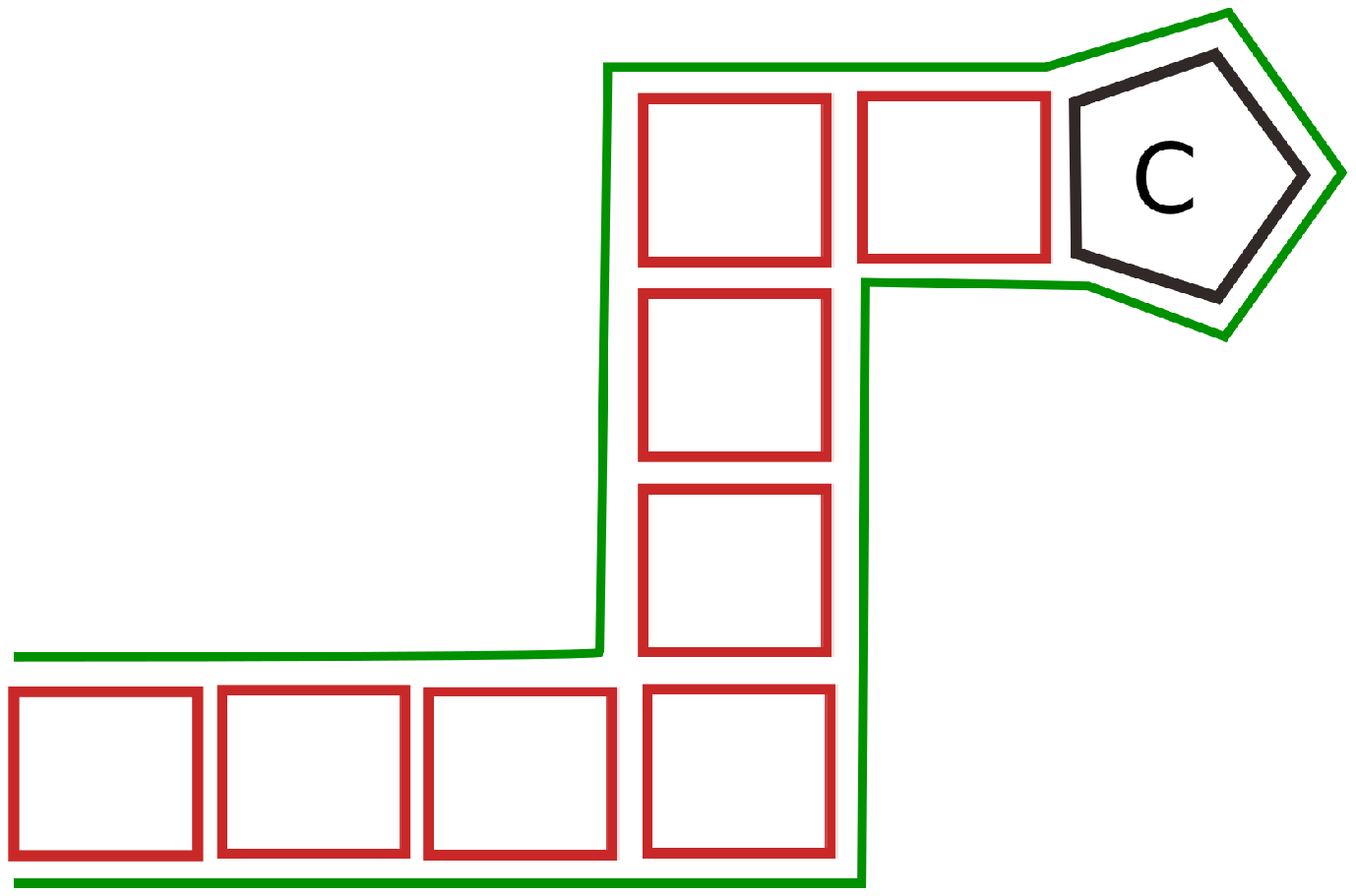}
\label{ribbon}
}
 \subfigure[~] 
 {
  \includegraphics[scale=0.3]{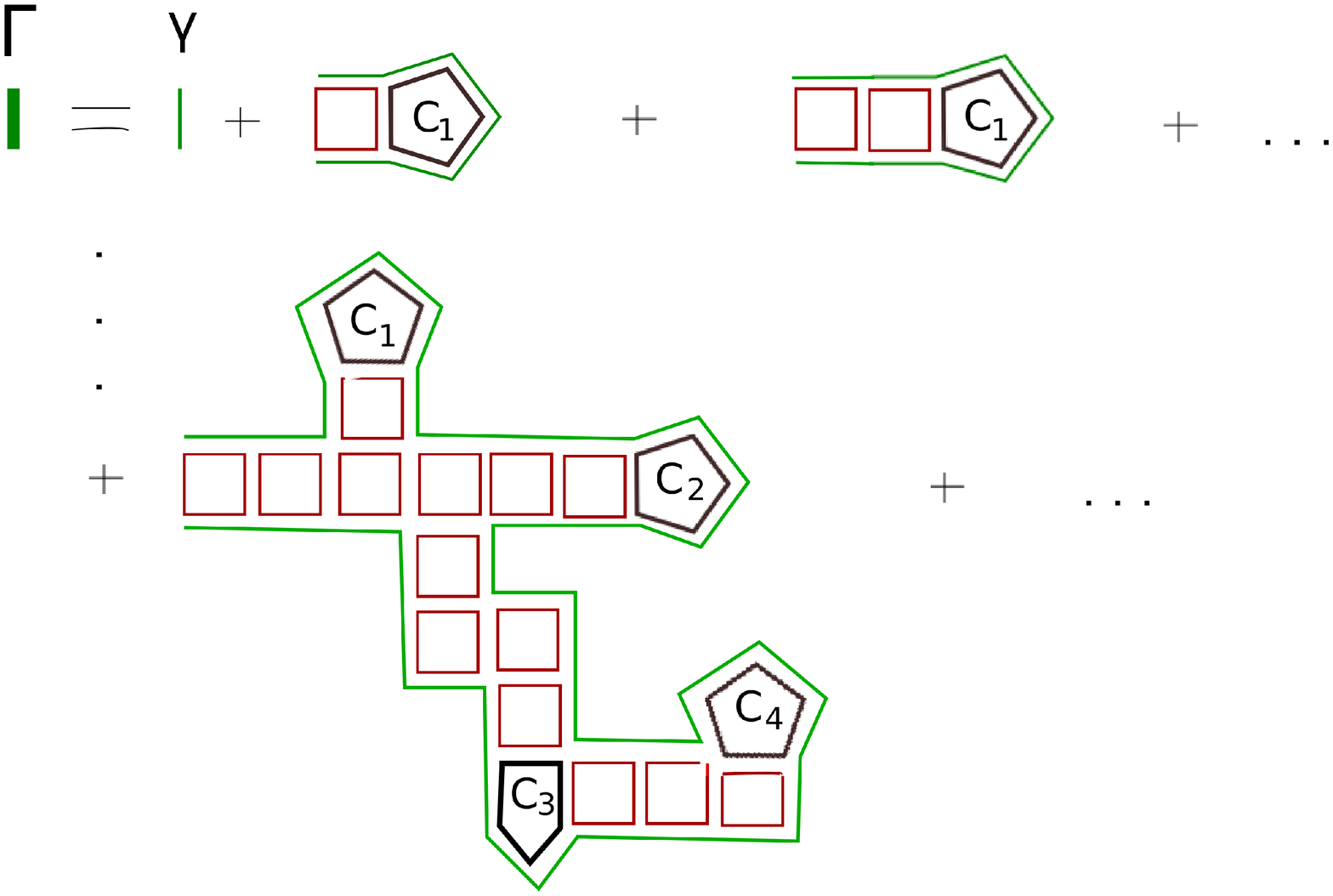}
\label{ribbon_sum}
}  
\caption{ (a) A ``ribbon'' diagram.  This is a chain, one lattice spacing in width, consisting of a series of plaquettes (red) taken from the expansion of the Wilson action, bordered by links taken from the expansion of the Higgs action (green), and terminating in a cluster $C$ (black).   (b) Schematic representation of a sample of diagrams which are implicitly summed in eq.\ \rf{dressG}.}
\end{figure*}

     Let us define a ``ribbon'' to be a one-plaquette wide strip of plaquettes bounded by Higgs lines on either side, both originating from an
expansion of the action in powers of $\b,\g$, beginning at a link $l$, and
terminating on a cluster, as shown in Fig.\ \ref{ribbon}.  Keeping $l$ fixed, we consider summing over all ribbons, and for each ribbon summing over the clusters at the endpoint.  Assuming for simplicity that the Higgs lines and plaquettes are in the fundamental representation (and this is not an important restriction, since the sum over representations is rapidly convergent), and that the ribbon is of length $d$, the ribbon is
associated (after integration over gauge and Higgs fields) with a weight
\beq
           \left({\b\over 4}\right)^{d}   \left({\g\over 4}\right)^{2d} 3 F(\C) \ ,
\eeq
where $F(\C)$ is the contribution from the cluster.  The factor of 3 comes from the fact that the cluster may be attached at either of the three
sides of the final plaquette in the ribbon.  Summing over all ribbons and all clusters we have
\bea
        Q(\b,\g) &<& \sum_{d=1}^\infty \left({\b\over 4}\right)^{d}   \left({\g\over 4}\right)^{2d}  9^d  \F(\b,\g) \non \\
                     &<&  {2^{-6}  9 \g^2 \b \over 1 - 2^{-6} 9 \g^2 \b} \F(\b,\g) \ ,
\label{Q}
\eea
where
\bea
\F(\b,\g) &=& 3 \sum_\C F(\C)  \ ,
\eea
and the inequality in \rf{Q} follows from the fact that there is some overcounting on the right hand side, since a self-avoiding constraint on the
sum has not been imposed.  The factor of $9^d$ derives from the fact that in building a ribbon plaquette by plaquette, then at the
{$n$-th} plaquette there are three links at which to join the $n+1$-th plaquette, which may be any one of the three plaquettes adjoining that link
which does not backtrack on the ribbon.   The smallest cluster is composed of four links.  If we sum up only $\g$-dependent contributions, then by previous methods we obtain an upper bound
\beq
    \F(\g~\mbox{only}) < 3  { (10\g)^{4} \over 1 - 10 \g} \ .
\eeq  
For small clusters there may also be significant $\b$ contributions, but these can be neglected for large clusters in which plaquettes would have to span large areas.

   Now define the ``dressed link'' factor $\G$ to be the solution of
\beq
           \G = \g + Q(\b,\G) \ .
\label{dressG}
\eeq
Diagrammatically, $\G$ is a sum of all tree-like structures connecting disconnected clusters, as indicated 
schematically in Fig.\ \ref{ribbon_sum}.
A bound on the contributions to $r_C$ from such tree-like structures is obtained from the bound \rf{bnd2} by replacing the $\g$ factors that arise in the sum over initial states by the dressed link factor $\G$, i.e.
\bea
          \tg^{L(C)+2T} { (10\G)^{L_0} \over 1 - 10\G} \ .
\eea
The reason is that this replacement accounts for the sum over all tree-like arrangements of disconnected clusters which terminate on links
of the ``trunk'' $\C_{xy}$, which is also summed over.

    Disconnected diagrams can also be joined by tubelike structures, such as the one appearing in Fig.\ \ref{disconnect}.    An estimate, at
 $T=1$, of the sum of tubes at constant time leaving a given plaquette and terminating in a cluster is
 \bea
         P(\b,\g) &<& \sum_{d=1}^\infty \left({\b\over 4}\right)^{4d}  15^d  \F(\b,\g)  \non \\
                      &<&  15 \left({\b\over 4}\right)^4 {\F(\b,\g)\over 1 - 15 \left({\b\over 4})\right)^4} \ .
\eea
where again the right hand side is actually an overestimate, due to ignoring a self-avoidance constraint.
The factor of 15 is the number of ways that that a tube of a given length can be extended at one end by one cube, and convergence
requires $\b \ll 4/15^{1/4} \approx 2$.  The tube can be extended, without backtracking, at one of five plaquettes at the end of the tube, in one of three possible steps (counting backward/forward in one of the possible directions) orthogonal to the plaquette.  Then in complete analogy to  \rf{dressG} we define a ``dressed plaquette'' factor
\beq
              \overline{\b} = \b + P(\overline{\b},\g) \ .
\eeq
Taking account of both tube and ribbon structures, we obtain the simultaneous equations
\beq
          \G = \g + Q(\overline{\b},\G)   ~~~,~~~   \overline{\b} = \b + P(\overline{\b},\G) \ .
\label{dressBG}
\eeq

    The prescription for including tree diagrams joining disconnected clusters, in order to get a bound on the right hand side of \rf{rdis},
is to replace the couplings $\b,\g$ in matrix elements for a single cluster in the initial state by the dressed factors $\overline{\b},\G$. 
But in the end, for $\tg \ll \tb \ll 1$, the solution of \rf{dressBG} is simply $\G \approx \g, ~ \overline{\b} \approx \b$, and the previous upper bound
on $r_C$ is not much affected.

\subsubsection{$\b \ll \l \ll 1/10$}

    As $\b$ is reduced, the time interval required for string breaking is also reduced.  For $\g \ll 1/10$, however, $\Psi_V$ is always
S$_c$ confining.  It is sufficient to consider the limiting case of $\b \ra 0$.  For $V$ a simple Wilson line, we have from eq.\ \rf{simpleloop}
that
\bea
             E_V &=& - \log\left[ W(L,1) \right] \non \\
                     &=& (-2\log \tg^2) L -\log(2\tg^2) \ .
\eea
which is obviously bounded by a linear potential.  In the more general case, $M_T(\C_2,\C_1)=M_\g(\C_2,\C_1)$.  Then, using the
Gershgorin Theorem and \rf{bnd2}, the largest eigenvalue of $M_T$ for $T=1$ is bounded from above by
\beq
            \l_{max}(1)  <  2 \tg^{2L_0 + 2} + \tg^{L_0(C)+2} { (10 \g)^{L_0} \over 1 - 10 \g}
\eeq
and the logarithm gives a lower bound to the energy, which again increases linearly with separation $L_0$.  Beyond $T=1$ we have string breaking.   To refine the estimate of the string-breaking time, one may go to a time-asymmetric lattice, with the lattice spacing $a_t$ in the
time direction much smaller than the lattice spacing $a$ in the space directions.  This asymmetry is accompanied by an increase in
the lattice coupling $\b_t$ associated with the timelike plaquettes.  Eventually $\b_t$ exceeds $\g$, and the preceding analysis can be
applied.

\subsubsection{Summary}

       The strong-coupling argument presented in this section is a bit lengthy, and despite the length it does not rise to the level of rigor required of a formal proof.  But the central idea is simple, and it boils downs
to this:  In the absence of matter loops, the energies of the $\Psi_V$ states are given by pure gauge theory, which we already know to be S$_c$ confining.  Inclusion of matter loops will eventually cause string breaking and a consequential loss of the linear potential in the course of Euclidean time evolution, but this event occurs only after the system has evolved for some finite time period.
If the strong-coupling conditions \rf{conditions} are satisfied and the Euclidean time $T$ obeys the bound \rf{T}, then the multiplicity
of screening contributions is outweighed by their exponential suppression in powers of $\g$, and the energy of a time-evolved
$\Psi_V$ state, obtained from the lattice logarithmic time derivative, is approximately that of the pure gauge theory.  Even for $\b \ll \g \ll 1/10$,
the S$_c$ condition is satisfied at $T=1$.
Hence S$_c$-confinement exists in some region of the $\b-\g$ phase diagram, and given the known result \cite{Greensite:2017ajx} that S$_c$-confinement does not exist everywhere in the phase plane, it follows that there is somewhere a transition line between the stronger separation-of-charge property in the confinement region and the weaker color-neutrality property in the Higgs region. 

\bigskip  \bigskip

\section{Symmetry Breaking and the \\ S$_c$-to-C transition}

    We conjecture that the transition from S$_c$ to C confinement coincides with the gauge-invariant symmetry breaking transitions seen
in Figs.\ \ref{phase} and \ref{su3gb}. The first question to ask is whether existing data on the location of C confinement, in the SU(2) gauge-Higgs theory, already rule this out.  

     In ref.\ \cite{Greensite:2017ajx} we considered three possible choices of $\Psi_V$ states:  the Dirac state (a non-abelian generalization of
charged states in an abelian theory), a ``pseudomatter'' state based on eigenmodes of the covariant Laplacian operator,
and a ``fat link'' Wilson line state derived from a familiar method of noise reduction in lattice gauge theory.  An S$_c$-to-C confinement transition was found for the first two states, but not for the third, which was
everywhere S$_c$ confining.  But it must be understood that if a region is S$_c$ confining, this behavior must be obtained not just for one
choice of $\Psi_V$ state, but for {\it all} such states.  In other words, S$_c$ confining behavior in a particular  $\Psi_V$ (e.g.\ the fat link state) in some region is a necessary but not sufficient condition for S$_c$ confinement in that region.  If even one $\Psi_V$ exhibits C confining behavior in a region, then that region is C confining.  Put another way, C confining behavior found for one state $\Psi_V$ in some region is a sufficient but not a necessary condition for C confinement in the region.  

   The transition from S$_c$ to C confinement in the Dirac state corresponds, as explained in ref.\ \cite{Greensite:2017ajx}, to the spontaneous breaking of a remnant gauge symmetry, global on each time slice, that exists in Coulomb gauge, and the location of that remnant symmetry breaking in SU(2) gauge-Higgs theory was found in ref.\ \cite{Caudy:2007sf}.  It is certain that the region in the $\b-\g$ phase diagram above the remnant symmetry breaking line is C confining.  But whether the region below this line is S$_c$ confining is uncertain, at least until
we come to the region of strong couplings, where the analysis of the previous section shows the existence of S$_c$ confinement.  

\begin{figure}[t!]
 \includegraphics[scale=0.7]{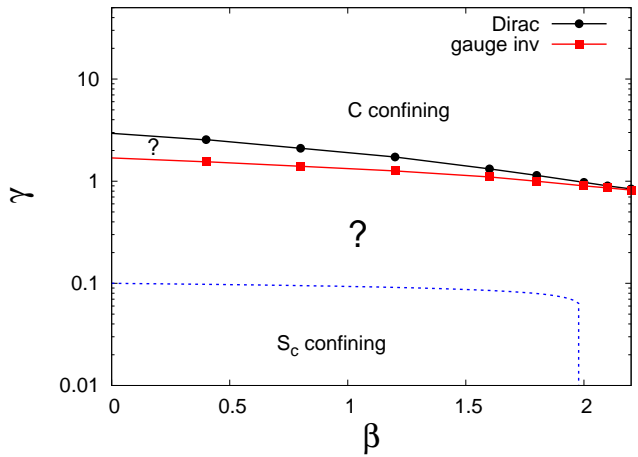}
\caption{For SU(2) gauge-Higgs theory, C confinement exists above the line denoted ``Dirac,'' and S$_c$-confinement exists in a 
strong-coupling region, as well as
along the line at $\b=0$.  The location of C and S$_c$ confinement in the rest of the phase diagram is uncertain.  Our conjecture
is that the S$_c$-to-C confinement transition line coincides with the gauge-invariant symmetry breaking line, denoted ``gauge inv''
in the figure.  Error bars on data points are, on this scale, smaller than the symbol size. } 
\label{sc}
\end{figure} 

   The situation at the moment is illustrated in Fig.\ \ref{sc}.  C confinement is known to exist above the Dirac line shown, but we do not know
how far it extends below that line.  S$_c$ confinement exists inside a strong-coupling region, whose boundary is indicated somewhat schematically in Fig.\ \ref{sc}, but we do not know how far it extends outside the region of convergence of the strong-coupling expansion. If one can find a $V$ operator such that $\Psi_V$ has C confining behavior anywhere below the
gauge-invariant symmetry breaking line, also displayed in Fig.\ \ref{sc}, then our conjecture about the coincidence of the  S$_c$-to-C and symmetry breaking transitions is wrong.  Only two points for the pseudomatter
transition were obtained in \cite{Greensite:2017ajx}, and one of these points (at $\b=1.2$) lies at a gamma value which is slightly below
the corresponding Dirac operator transition.\footnote{The other point, at $\b=2.2$, coincides with both the Dirac and gauge-invariant transitions.}  That point is still above the gauge invariant transition, however.  This means that at least some
of the C confining region lies below the Dirac transition, and the conjecture is that the entire region between the Dirac line and the gauge invariant transition line is C confining, while the region below the gauge invariant transition line is S$_c$ confining.  

   So the existing data is at least consistent with our conjecture.  To proceed further, some effort must be devoted to inventing and testing more operators which might falsify (or, alternatively, support) this proposal.   A first step would be to test operators, already studied in \cite{Greensite:2017ajx} for SU(2) gauge-Higgs theory, in the SU(3) gauge-Higgs case.  We hope to report on these efforts at a later time.

\section{Conclusions}

    There exist global symmetries in the Higgs sector of gauge-Higgs theories which are independent of
any gauge choice, and these symmetries can break spontaneously in the sense explained in section II, where we also explain the absence of Goldstone excitations in the full theory.  We have constructed gauge-invariant order parameters which can detect  the spontaneous breaking of such symmetries.  There are two obvious questions, both relating to the nature of this transition.  First, given the result of Osterwalder and Seiler \cite{Osterwalder:1977pc}, this transition cannot correspond
to a thermodynamic transition everywhere along the transition line.  So is it possible to speak of a phase transition which does not correspond
to a non-analyticity in the free energy?  In fact there are examples of such transitions, namely the Kertesz transition line \cite{Kertesz}
found in Ising and Potts models in an external magnetic field.\footnote{The possible relevance of this example in the context of pure gauge theories at finite temperature, and in gauge-Higgs models, has been discussed by a number of authors 
\cite{Langfeld:2002ic,Fortunato:1999wr,*Wenzel:2005nd,*Bertle:2003pj}.}
But the next question is what is the physical difference between the symmetric
and broken phases in a gauge-Higgs theory.  If there is no physical difference and no singular behavior in the free energy, then this
transition is physically meaningless.  However, we believe there is a natural candidate for the physical difference between the two phases, and that is the distinction between separation-of-charge (S$_c$) confinement and color (C) confinement.    In section VI 
we have shown that S$_c$ confinement must exist somewhere in the $\b-\g$ coupling plane, and given the fact  \cite{Greensite:2017ajx}  that 
S$_c$ confinement does not exist throughout the plane, there must be a transition between these physically distinct phases.

   So we will conclude this article by repeating the conjecture, made in the previous section, that the gauge-invariant global symmetry breaking transition that we have located in gauge-Higgs theory coincides with the transition that must exist between the S$_c$ and C confinement.  If so, this transition separates two  phases that can be meaningfully distinguished as confinement vs.\ Higgs, in which a global symmetry  is either unbroken, or spontaneously broken in sense explained above.  
   
\acknowledgments{We thank Axel Maas for helpful discussions.  JG's research is supported by the U.S.\ Department of Energy under Grant No.\ DE-SC0013682.}   

\bibliography{sym3}

\end{document}